\documentclass[floatfix, superscriptaddress, twocolumn, showpacs, aps, pra, 10pt,amsmath,amssymb]{revtex4-2}
\usepackage{graphicx}
\usepackage{dcolumn}
\usepackage{bm}
\usepackage{stackrel}
\usepackage{amsmath}
\usepackage{amsfonts}
\usepackage{amssymb}
\usepackage{amsthm}
\usepackage{epstopdf}
\usepackage{longtable}
\usepackage{array}
\usepackage{pifont}
\usepackage{latexsym}
\usepackage[latin1]{inputenc}
\usepackage{bm}
\usepackage{textcomp}
\usepackage{subfigure}
\usepackage{braket}
\usepackage{wasysym}
\usepackage{ulem}
\usepackage{color, soul}
\usepackage{pifont}
\usepackage{multirow}

\begin{document}

\title{SU(2) hyper-clocks:\\ quantum engineering of spinor interferences for time and frequency metrology}
\author{T. Zanon-Willette}
\affiliation{Sorbonne Université, Observatoire de Paris, Université PSL, CNRS, LERMA, F-75005, Paris, France}
\email{thomas.zanon@sorbonne-universite.fr}
\affiliation{MajuLab, International Research Laboratory IRL 3654, Université Côte d'Azur, Sorbonne Université, National University of Singapore, Nanyang
Technological University, Singapore,}
\affiliation{Centre for Quantum Technologies, National University of Singapore, 117543 Singapore, Singapore}
\author{D. Wilkowski}
\address{MajuLab, International Research Laboratory IRL 3654, Université Côte d'Azur, Sorbonne Université, National University of Singapore, Nanyang
Technological University, Singapore,}
\address{Centre for Quantum Technologies, National University of Singapore, 117543 Singapore, Singapore}
\address{School of Physical and Mathematical Sciences, Nanyang Technological University, 637371 Singapore, Singapore}
\author{R. Lefevre}
\address{Department of Physics, Royal Holloway, University of London, Royal Holloway Egham Hill, Egham TW20 0EX, United Kingdom}
\author{A.V. Taichenachev}
\address{Novosibirsk State University, ul. Pirogova 2, 630090 Novosibirsk, Russia}
\address{Institute of Laser Physics, Siberian Branch, Russian Academy of Sciences, prosp. Akad. Lavrent'eva 15B, 630090 Novosibirsk, Russia}
\author{V.I. Yudin}
\address{Novosibirsk State University, ul. Pirogova 2, 630090 Novosibirsk, Russia}
\address{Institute of Laser Physics, Siberian Branch, Russian Academy of Sciences, prosp. Akad. Lavrent'eva 15B, 630090 Novosibirsk, Russia}
\address{Novosibirsk State Technical University, prosp. Karla Marksa 20, 630073 Novosibirsk, Russia}

\begin{abstract}
In 1949, Ramsey's method of separated oscillating fields was elaborated boosting over many decades metrological performances of atomic clocks and becoming the standard technique for very high precision spectroscopic measurements. A generalization of this interferometric method is presented replacing the two single coherent excitations by arbitrary composite laser pulses. The rotation of the state vector of a two-level system under the effect of a single pulse is described using the Pauli matrices basis of the SU(2) group. It is then generalized to multiple excitation pulses by a recursive Euler-Rodrigues-Gibbs algorithm describing a composition of rotations with different rotation axes. A general analytical formula for the phase-shift associated with the clock's interferometric signal is derived. As illustrations,
hyper-clocks based on three-pulse and five-pulse interrogation protocols are studied and shown to exhibit nonlinear cubic and quintic sensitivities to residual probe-induced light-shifts. The presented formalism is well suited to optimize composite phase-shifts produced by tailored quantum algorithms in order to design a new generation of optical frequency standards and robust engineering control of atomic interferences in AMO physics with cold matter and anti-matter.
\end{abstract}

\date{\today}

\preprint{APS/123-QED}

\maketitle

\section{Introduction}

\indent The method of separated oscillating fields was introduced by Ramsey in 1949 to improve frequency resolution of spectroscopic measurements and collect information about the internal structure of atoms and molecules~\cite{Ramsey:1949,Ramsey:1950,Ramsey:1956}. Today, understanding how to improve the robustness of spectroscopy with coherent radiation by reducing or eliminating laser probe-induced systematics still remains a central goal in the broad and important field of robust atomic sensors from stringent tests of fundamental physics to quantum metrology with optical clocks and matter-wave interferometry~\cite{Zanon-Willette:2022}.

Ramsey derived in 1950 the first original quantum mechanical description of a spin 1/2 interferometric resonance with two separated coherent pulses by using a Schrödinger wave-function description~\cite{Ramsey:1950} later extending the analysis to phase jump, pulse shapes and amplitudes~\cite{Ramsey:1951,Ramsey:1958}. The Ramsey's method became the standard technique in atomic physics based on microwave and laser spectroscopy and in quantum metrology with atomic beams~\cite{Essen:1955} and cold atomic fountains~\cite{Clairon:1991} to measure transition frequencies between particle states with very high-precision~\cite{Ramsey:1990}. After 70 years, Ramsey interferometry is still a powerful tool to investigate matter-light interaction with a few particles such as in modern cavity QED experiments on Schr\"{o}dinger's cats with Rydberg's atoms~\cite{Haroche:2006,Gleyzes:2007,Haroche:2013}, in quantum information with trapped ions~\cite{McCormick:2019} or with superconducting qubits~\cite{Krantz:2019}.
\begin{figure}[t!!]
\center
\resizebox{8.5cm}{!}{\includegraphics[angle=0]{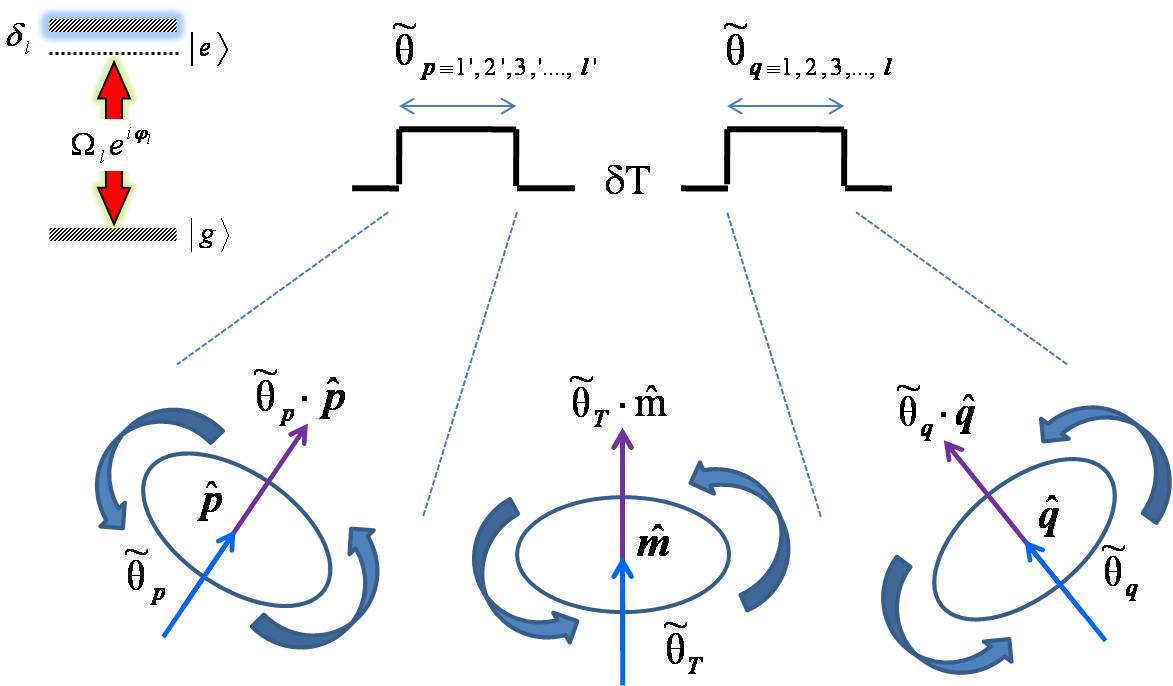}}
\caption{(color online) Generalized Ramsey's method of separated oscillating fields with angle-axis representation. Single pulses (or sets of composite pulses), around a single Ramsey free evolution phase as $\delta\textup{T}$, are introduced by an effective pulse area $\widetilde{\theta}_{p}$ and $\widetilde{\theta}_{q}$ where $p\equiv 1',2',3'...$, $q\equiv 1,2,3...$ with arbitrary rotation axis orientation $\widehat{n}_{p}$ and $\widehat{n}_{q}$ around the $\widehat{m}$ axis. Single or composite laser pulse parameters are including laser phase $\varphi_{p},\varphi_{q}$, field excitation $\Omega_{p},\Omega_{q}$, pulse duration $\tau_{p},\tau_{q}$ and a frequency detuning $\delta_{p}=\delta\mp\Delta_{p},\delta_{q}=\delta\mp\Delta_{q}$ including the uncompensated part of the light-shift $\Delta_{p},\Delta_{q}$ during pulses~\cite{Yudin:2009}. Any residual light-shift is inducing a weak distortion of the angle-axis orientation relatively to the intermediate free evolution zone.}
\label{fig:fig1}
\end{figure}

Nevertheless, the architecture of the two-pulse Ramsey interferometer as shown in Fig.~\ref{fig:fig1} has remained untouched until 2010 when an improvement for clock spectroscopy was proposed~\cite{Yudin:2009,Yudin:2010} and experimentally applied to a single trapped ion~\cite{Huntemann:2012}. A Ramsey sequence of two coherent laser pulses is used with a pre-compensation of the estimated light-shift including a third intermediate pulse which is inserted to act like a spin-echo compensation of field amplitude error~\cite{Yudin:2009,Yudin:2010,Zanon-Willette:2015,Taichenachev:2017}.
After the experimental success of the hyper-Ramsey protocol to drastically reduce, by four orders of magnitude, the residual light-shift on the single-ion $^{171}$Yb$^{+}$ octupole clock transition~\cite{Huntemann:2016}, more robust generalized hyper-Ramsey three-pulse protocols have been discovered against laser pulses-induced frequency-shifts~\cite{Zanon-Willette:2018} including decoherence effect~\cite{Tabatchikova:2013,Yudin:2016,Zanon-Willette:2017}.
Naturally, the question arises if there is a way to extend interrogation protocols to any sets of arbitrary composite pulses around a single free evolution time in a symmetrical fashion.
The first positive answer including composite laser pulses was provided through a Cayley-Klein parametrization of rotation spinors~\cite{Zanon-Willette:2019}. An extended model based on multiple hyper-Ramsey-Bordé building-blocks with two-level operators and quantization of motion has been recently developed where arbitrary composite optical pulses are used not only to shield quantum clock interferences against residual light-shift but also to protect atomic matter-waves against laser probe-induced frequency-shifts at ultra-cold temperature~\cite{Zanon-Willette:2022}.
A complementary approach to~\cite{Zanon-Willette:2022}, extending~\cite{Zanon-Willette:2019}, would be strongly helpful in designing new sequences of laser pulses to compensate for their systematic errors in pulse detuning, relative phase and/or pulse area through various robust quantum control techniques~\cite{Vandersypen:2004,Jones:2011,Merrill:2014,Torosov:2019,Torosov:2022}.

\indent The purpose of this work is to present an alternative formalism to refs~\cite{Yudin:2010,Zanon-Willette:2019,Zanon-Willette:2022} allowing a convenient derivation of generalized hyper-Ramsey clock interferences and atomic phase-shifts with arbitrary composite pulses and axis orientation. Exponentials of Pauli matrices~\cite{Pauli:1927} are used to decompose a complex transition amplitude following an initial suggestion by I.I. Rabi, N.F. Ramsey and J. Schwinger~\cite{Rabi:1954}. Multiple interactions between the two-level system (or qubit) and laser pulses will be treated as a composition of spinor rotations on a Bloch-sphere.
Further, a recursive algorithm based on the Euler-Rodrigues-Gibbs geometrical transformation for dual axis rotation composition~\cite{Euler:1770,Rodrigues:1840,Gibbs:1901,Dai:2015,Valdenebro:2016} is extended to qubit rotation allowing a systematic exploration and optimization of more elaborated interrogation protocols with multiple laser pulses.

The recursive algorithm has been inspired by composite pulses developed originally in Nuclear Magnetic Resonance (NMR)~\cite{Levitt:1986,Siminovitch-I:1997,Siminovitch-II:1997} where composition of two rotations with quaternion computation rules~\cite{Altmann:1986,Blumich:1985} have already been applied to facilitate geometrical analysis and role of symmetry in the design of composite pulse action on nuclear spins ensemble~\cite{Councell:1985,Levitt:2008}.
More recently, the Schrödinger equation has been re-explored within a quaternionic representation of Pauli spinor of an electron~\cite{Cahay:2019} and a quaternionic derivation of the Ramsey transition probability has been presented~\cite{Cahay:2020} providing an alternative way to compute composite rotations on the Bloch-sphere.

The paper is organized as follow: in section II, we introduce a vectorial representation of spinor matrix components associated to complex transition amplitudes. Then, a compact expression of the composite phase-shift associated to quantum interferences with multi-pulses is given in section III. From a quantum engineering perspective, the best tailoring approach of atomic interferences is tracked to produce, by pulse engineering methods such as quantum control~\cite{Schirmer:2002,Glaser:2015}, an optimization of some targeted performances, i.e frequency-shift and signal amplitude of optical clocks, make more robust to important variations of relevant experimental parameters.
Finally, exact expressions of atomic phase-shifts are derived. Here, we are mainly focusing on specific laser pulse protocols on three-pulse and five-pulse schemes related to the design of hyper-Ramsey composite phase-shifts for ultra-robust optical clocks~\cite{Yudin:2010,Zanon-Willette:2016,Zanon-Willette:2019}.
Such hyper-clocks produce various highly nonlinear, flexible and robust compensation of the residual light-shift with a different sensitivity to laser probe intensity fluctuation~\cite{Beloy:2018}.
SU(2) hyper-clocks are a class of optical qubit-clocks based on composite laser pulse protocols aim at reducing laser probe-induced frequency shifts by several order of magnitude improving the accuracy of optical clocks.

\section{Vectorial representation of generalized hyper-Ramsey spinor components}

\subsection{Pauli-spin decomposition}

\indent The model is based on a SU(2) Pauli-spin decomposition of generalized hyper-Ramsey resonances and phase-shifts.
Exact expressions are derived for spinor components of a unitary interaction matrix describing coherent interaction between a qubit and laser excitation pulses.
The time-dependent atomic wave-function $\Psi(t)=C_{g}(t)|g\rangle+C_{e}(t)|e\rangle$ is interacting with two pulses labeled by $p$ and $q$ and separated by a single free evolution time T as reported in Fig.~\ref{fig:fig1}, inducing a qubit rotation composition as~\cite{Rabi:1954}:
\begin{equation}
\begin{split}
\Psi(t)&=e^{i\widetilde{\theta}_{q}(\widehat{n}_{q}\cdot\overrightarrow{\sigma})}e^{i\theta_{m}(\widehat{m}\cdot\overrightarrow{\sigma})}e^{i\widetilde{\theta}_{p}(\widehat{n}_{p}\cdot\overrightarrow{\sigma})}\cdot\Psi(0)\\
&=_{p}^{q}\textup{C}\cdot\Psi(0)
\end{split}
\label{eq:rotation}
\end{equation}
The Pauli vector is defined by $\overrightarrow{\sigma}=\sigma_{x}\widehat{x}+\sigma_{y}\widehat{y}+\sigma_{z}\widehat{z}$.
Rotation axis definitions corresponding to Eq.~\ref{eq:rotation} are introduced by $\widehat{n}_{p}=\overrightarrow{n}_{p}/\|\overrightarrow{n}_{p}\|$, $\widehat{m}=\overrightarrow{m}/\|\overrightarrow{m}\|$ and $\widehat{n}_{q}=\overrightarrow{n}_{q}/\|\overrightarrow{n}_{q}\|$. Rotation angles and angular velocities are defined by $\widetilde{\theta}_{p}=\|\overrightarrow{n}_{p}\|\tau/2$, $\widetilde{\theta}_{m}=\|\overrightarrow{m}\|\textup{T}/2$ and $\widetilde{\theta}_{q}=\|\overrightarrow{n}_{q}\|\tau/2$, with cartesian unit vector coordinates $\overrightarrow{n}_{p}=(n_{p_{x}},n_{p_{y}},n_{p_{z}})$, $\overrightarrow{m}=(m_{x},m_{y},m_{z})$ and $\overrightarrow{n}_{q}=(n_{q_{x}},n_{q_{y}},n_{q_{z}})$~\cite{Siminovitch-I:1997,Siminovitch-II:1997}.
The $2\times2$ matrix components $_{p}^{q}C_{u,u'}$ are written as:
\begin{equation}
\begin{split}
_{p}^{q}\textup{C}=\left(\begin{array}{cc}
            _{p}^{q}C_{gg} & _{p}^{q}C_{ge} \\
             _{p}^{q}C_{eg} & _{p}^{q}C_{ee} \\
          \end{array}
        \right)
\end{split}
\end{equation}
where $u,u'=g,e$.
Relations between the components of the unitary interaction matrix correspond to the SU(2) group, namely $_{p}^{q}C_{gg}=\{_{p}^{q}C_{ee}\}^{*}$, $_{p}^{q}C_{ge}=-\{_{p}^{q}C_{eg}\}^{*}$, $|_{p}^{q}C_{gg}|^{2}+|_{p}^{q}C_{ge}|^{2}=1$.

Any general unitary operator corresponding to a rotation of a qubit around a rotation axis $\widehat{n}_{l}$ with a rotation angle $\widetilde{\vartheta}_{l}$ ($l=p,q$) is evaluated by the exponential Pauli-spin decomposition~\cite{Pauli:1927,Yepez:2013}:
\begin{equation}
e^{i\widetilde{\vartheta}_{l}(\widehat{n}_{l}\cdot\overrightarrow{\sigma})}=\sigma_{0}\cos\widetilde{\vartheta}_{l}+i(\widehat{n}_{l}\cdot\overrightarrow{\sigma})\sin\widetilde{\vartheta}_{l}
	\label{eq:spinor}
\end{equation}
with the identity Pauli matrix $\sigma_{0}$ and satisfying the vectorial identity relation~\cite{Yepez:2013}:
\begin{equation}
(\widehat{n_{p}}\cdot\overrightarrow{\sigma})\cdot(\widehat{n}_{q}\cdot\overrightarrow{\sigma})=(\widehat{n}_{p}\cdot\widehat{n}_{q})\sigma_{0}+i(\widehat{n}_{p}\times\widehat{n}_{q})\cdot\overrightarrow{\sigma}
\end{equation}
The computational procedure calculates the transition probability for a reorientation of the qubit state $u$ into a state $u'$. It is simply given by $_{p}^{q}P_{uu'}=|_{p}^{q}C_{uu'}|^{2}$ where each spinor component of the unitary matrix is expressed as~\cite{Zanon-Willette:2015,Zanon-Willette:2018,Zanon-Willette:2017,Zanon-Willette:2019}:
\begin{equation}
\begin{split}
_{p}^{q}C_{uu'}=_{p}^{q}\widetilde{C}^{+}_{uu'}e^{i_{p}^{q}\widetilde{\Phi}_{uu'}^{+}}e^{i\widetilde{\theta}_{m}}+_{p}^{q}\widetilde{C}^{-}_{uu'}e^{i_{p}^{q}\widetilde{\Phi}_{uu'}^{-}}e^{-i\widetilde{\theta}_{m}}
\end{split}
\label{eq:transition-probability-Pauli}
\end{equation}
The phase-shift difference between components $_{p}^{q}\widetilde{C}^{\pm}_{uu'}$ is introduced as:
\begin{equation}
\begin{split}
_{p}^{q}\widetilde{\Phi}_{uu'}&=_{p}^{q}\widetilde{\Phi}_{uu'}^{+}-_{p}^{q}\widetilde{\Phi}_{uu'}^{-}
\label{eq:general-phase-shift}
\end{split}
\end{equation}
Envelopes $_{p}^{q}\widetilde{C}_{uu'}^{\pm}$ are themselves expressed with a complex modulus as~\cite{Abramowitz:1968,Sangwine:2010}:
\begin{equation}
\begin{split}
_{p}^{q}\widetilde{C}_{uu'}^{\pm}=\frac{1}{2}\left(\cos\widetilde{\theta}_{p}\cos\widetilde{\theta}_{q}\right)~_{p}^{q}C^{\pm}\sqrt{1+\tan^{2}\left(_{p}^{q}\widetilde{\Phi}_{uu'}^{\pm}\right)}
\label{eq:envelop-1}
\end{split}
\end{equation}
and
\begin{equation}
\begin{split}
_{p}^{q}C^{\pm}=&\sigma_{0}\pm\widehat{m}\cdot\overrightarrow{\sigma}\pm\left[_{p}^{q}\widehat{N}_{-}\times\widehat{m}\right]\cdot\overrightarrow{\sigma}-~_{p}^{q}\widehat{N}^{\widehat{m}}_{\bullet}
\label{eq:envelop-2}
\end{split}
\end{equation}
Atomic phase-shifts $\widetilde{\Phi}_{uu'}(\pm)$ associated to $_{p}^{q}\widetilde{C}^{\pm}_{uu'}$ are also evaluated with Pauli-spin matrices and are expressed with a complex argument~\cite{Abramowitz:1968,Sangwine:2010}:
\begin{equation}
\tan_{p}^{q}\widetilde{\Phi}_{uu'}^{\pm}\equiv\frac{_{p}^{q}\widehat{N}_{+}\cdot\left[\overrightarrow{\sigma}\pm\widehat{m}\sigma_{0}\right]+~_{p}^{q}\widehat{N}_{\times}\cdot\left[\overrightarrow{\sigma}\mp\widehat{m}\sigma_{0}\right]}
{\sigma_{0}\pm\widehat{m}\cdot\overrightarrow{\sigma}\pm\left[_{p}^{q}\widehat{N}_{-}\times\widehat{m}\right]\cdot\overrightarrow{\sigma}-~_{p}^{q}\widehat{N}^{\widehat{m}}_{\bullet}}
\label{eq:complex-phase-shift}
\end{equation}
where:
\begin{equation}
\begin{split}
_{p}^{q}\widehat{N}_{+}&\equiv\widehat{n}_{p}\tan\widetilde{\theta}_{p}+\widehat{n}_{q}\tan\widetilde{\theta}_{q}\\
_{p}^{q}\widehat{N}_{-}&\equiv\widehat{n}_{p}\tan\widetilde{\theta}_{p}-\widehat{n}_{q}\tan\widetilde{\theta}_{q}\\
_{p}^{q}\widehat{N}_{\times}&\equiv\widehat{n}_{p}\tan\widetilde{\theta}_{p}\times\widehat{n}_{q}\tan\widetilde{\theta}_{q}\\
_{p}^{q}\widehat{N}^{\widehat{m}}_{\bullet}&\equiv\left(\widehat{n}_{p}\cdot\widehat{n}_{q}\right)_{\widehat{m},\overrightarrow{\sigma}}\tan\widetilde{\theta}_{p}\tan\widetilde{\theta}_{q}
\label{eq:reduced-variables-1}
\end{split}
\end{equation}
with a reduced variable:
\begin{equation}
\begin{split}
\left(\widehat{n}_{p}\cdot\widehat{n}_{q}\right)_{\widehat{m},\overrightarrow{\sigma}}=&\left(\sigma_{0}\mp\widehat{m}\cdot\overrightarrow{\sigma}\right)\left(\widehat{n}_{p}\cdot\widehat{n}_{q}\right)\\
&\pm\left[\left(\widehat{m}\cdot\widehat{n}_{p}\right)\widehat{n}_{q}+\left(\widehat{m}\cdot\widehat{n}_{q}\right)\widehat{n}_{p}\right]\cdot\overrightarrow{\sigma}
\label{eq:reduced-variables-2}
\end{split}
\end{equation}
Frequency shifts of atomic interferences produced by the laser probe excitation scheme are described by Eq.~\ref{eq:complex-phase-shift} where the influence of the light-shift is to change simultaneously the rotation axis orientation and the effective Rabi frequency as shown in Fig~\ref{fig:fig1}.
This equation contains a dot-product (scalar) term as $\widehat{N}_{\bullet}$ and a cross-product (vectorial) term as $\widehat{N}_{\times}$ that are effectively related to a composition rule of two unit quaternions~\cite{Siminovitch-I:1997,Altmann:1986,Blumich:1985} and to the Euler-Rodrigues-Gibbs (ERG) formula for 3D rotation composition~\cite{Euler:1770,Rodrigues:1840,Gibbs:1901,Valdenebro:2016}.
Pauli-spin matrices $\sigma_{x,y,z}$ as well as the identity matrice $\sigma_{0}$ are used as Hilbert-space pointers to individually address each $_{p}^{q}\widetilde{\Phi}_{uu'}^{\pm}$ component associated to diagonal and off-diagonal elements of the spinor matrix~\cite{Feynman:1982}.
All $_{p}^{q}C_{uu'}$ components of a rotated qubit by Ramsey spectroscopy with composite pulses can be analytically derived using the Pauli-spin model presented above.
\begin{figure}[t!!]
\center
\resizebox{7.5cm}{!}{\includegraphics[angle=0]{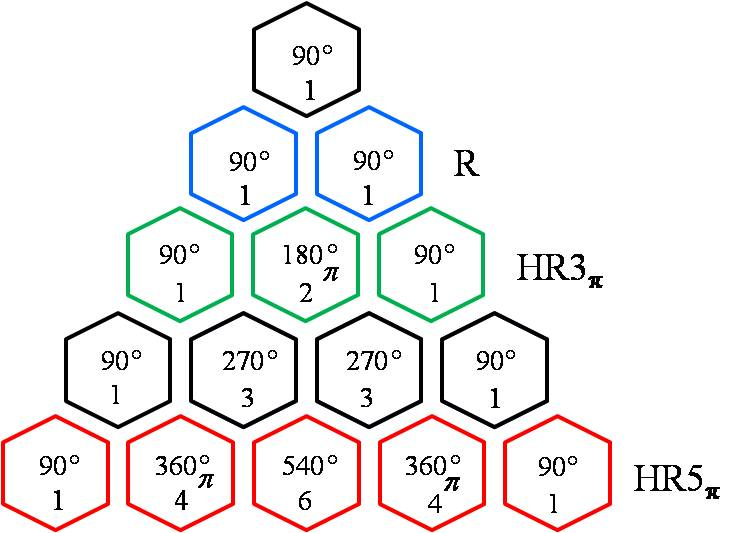}}
\caption{(color online) Hyper-clock protocol classification following the Pascal's triangle for binomial coefficients representation. Appropriate laser phase-jumps of $\pi$ are inserted as subscripts within specific laser pulses. Pulse areas are here indicated in integer units of a $90^{\circ}$ reference pulse following the time length of each single pulse. The qubit free-rotation denoted $\delta\textup{T}$ is placed either between the first two pulses or between the last two pulses of HR3$_{\pi}$ and HR5$_{\pi}$ protocols.}
\label{fig:protocols}
\end{figure}

\subsection{Euler-Rodrigues-Gibbs (ERG) transformation rules and recursive algorithm}

\indent Turning to a generalized hyper-Ramsey resonance with an arbitrary number of composite pulses, left and right single Pauli-spin qubits from Fig.~\ref{fig:fig1} should be now replaced by composite qubits as following:
\begin{equation}
\begin{split}
e^{i\widetilde{\theta}_{p}(\widehat{n}_{p}\cdot\overrightarrow{\sigma})}&\equiv\overrightarrow{\prod}_{l=\textup{1'}}^{\textup{p}}e^{i\widetilde{\theta}_{l}(\widehat{n}_{l}\cdot\overrightarrow{\sigma})}\\
e^{i\widetilde{\theta}_{q}(\widehat{n}_{q}\cdot\overrightarrow{\sigma})}&\equiv\overleftarrow{\prod}_{l=\textup{1}}^{\textup{q}}e^{i\widetilde{\theta}_{l}(\widehat{n}_{l}\cdot\overrightarrow{\sigma})}
\end{split}
\label{eq:product}
\end{equation}
where each arrow indicates the direction to develop the multi-pulse product with growing indices.

In order to track analytically the resulting phase-shift associated to composite interferences, a recursive algorithm is presented based on iteration of the Euler-Rodrigues-Gibbs (ERG) transformation applied to composite pulses from left and right sides of the two-pulse interferometer. The ERG transformation rules, acting on unit vector coordinates, for a given set of $l\in\{p,q\}$ ($-$ for p and $+$ for q) composite pulses are given by:
\begin{eqnarray}
\left\{
\begin{split}
\cos\widetilde{\theta}_{l}&\mapsto\cos\widetilde{\theta}_{l}\cos\widetilde{\theta}_{l+1}\left[1-_{l}^{l+1}\widehat{N}^{0}_{\bullet}\right]\\
\widehat{n}_{l}\tan\widetilde{\theta}_{l}&\mapsto\frac{_{l}^{l+1}\widehat{N}_{+}\pm~_{l}^{l+1}\widehat{N}_{\times}}{1-_{l}^{l+1}\widehat{N}^{0}_{\bullet}}
\end{split}
\right.
\label{eq:transformation-pq-pulses}
\end{eqnarray}
with $_{l}^{l+1}\widehat{N}^{0}_{\bullet}\equiv\widehat{n}_{l}\cdot\widehat{n}_{l+1}\tan\widetilde{\theta}_{l}\tan\widetilde{\theta}_{l+1}$.

These rules applied on $_{p}^{q}C_{uu'}$ components, used as a quantum-processing algorithm, are iterated $p-1$ and $q-1$ times when running with an ensemble of $\{p,q\}$ pulses (see appendix for an example). A different recursive algorithm has been developed in~\cite{Zanon-Willette:2022} related to a Möbius transformation in conformal mapping~\cite{Lee:2002}, for instance see the reference note~\cite{Note}.
A complete geometrical representation of the qubit dynamics is achieved through Feynman-Vernon-Hellwarth coordinates to visualize composite rotations on a Bloch-sphere~\cite{Feynman:1957}.
A straightforward extension of generalized hyper-Ramsey resonances and phase-shifts to a higher quantum J spin made of composite qubits with equally energy spaced levels (hyper qudit-clock) is provided by application of the Majorana formula~\cite{Majorana:1932,Bloch:1945,Schwinger:1977} or by using a polynomial matrix expansion of spin rotation~\cite{Curtright:2014}.
\begin{figure}[t!!]
\center
\resizebox{8.5cm}{!}{\includegraphics[angle=0]{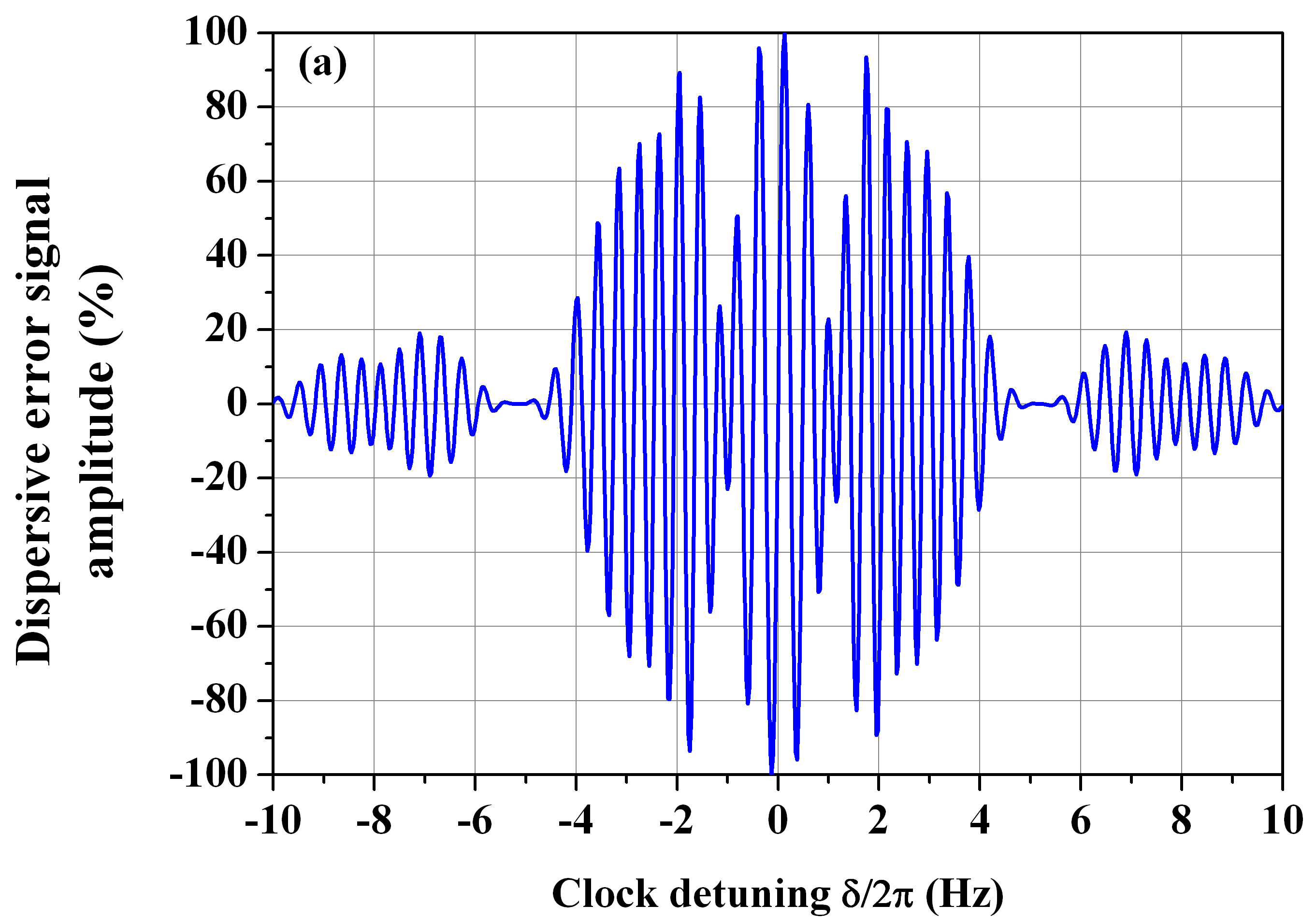}}
\resizebox{8.5cm}{!}{\includegraphics[angle=0]{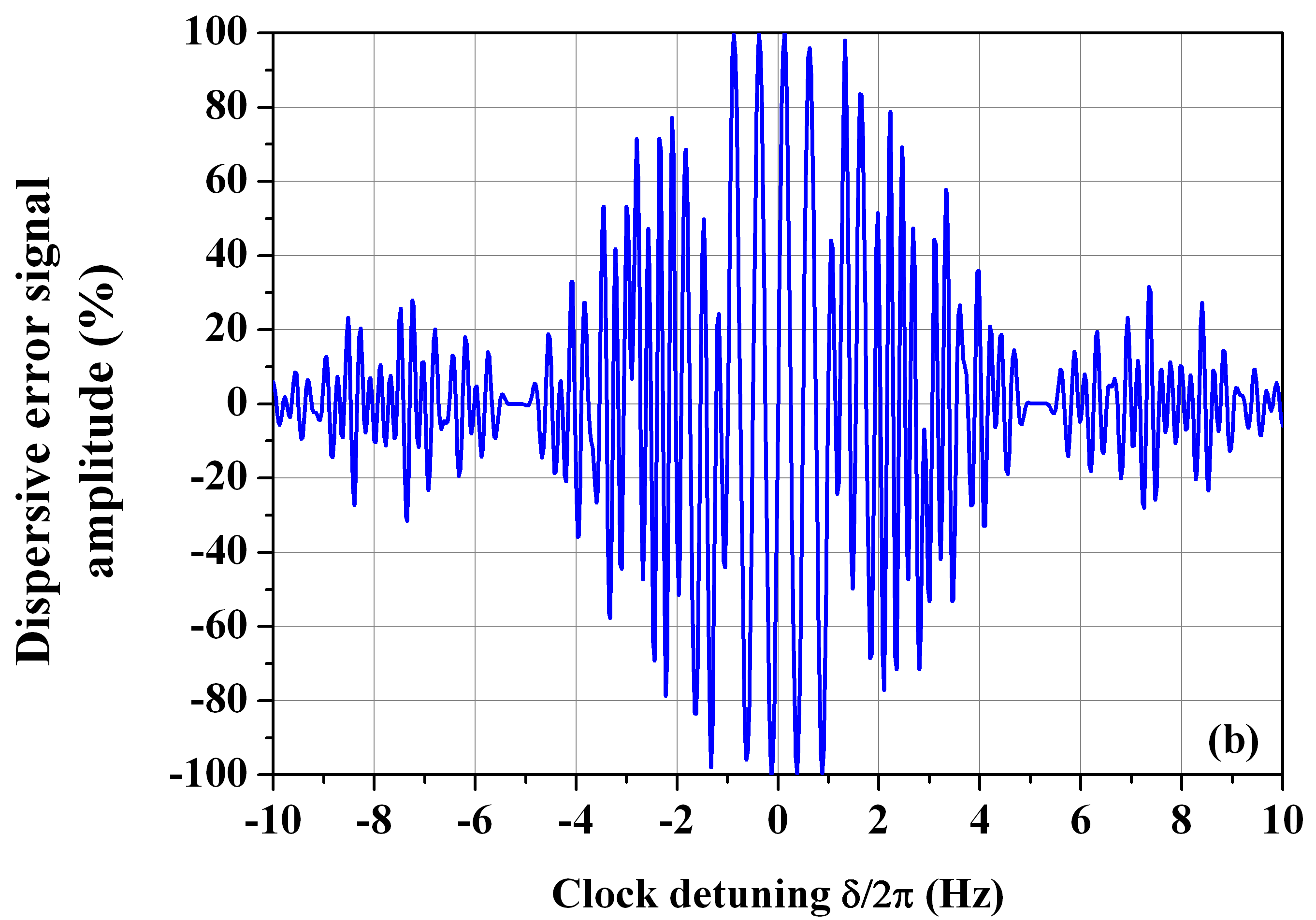}}
\caption{(color online). Two dispersive error signals, calculated from Eq.~\ref{eq:clock-interference}, are plotted versus the clock frequency detuning. (a) HR3$_{\pi}$ protocol as $90'^{\circ}_{\pm\pi/2}\dashv\delta\textup{T}\vdash180_{\pi}^{\circ}90^{\circ}$, (b) HR5$_{\pi}$ protocol as $90'^{\circ}_{\pm\pi/2}\dashv\delta\textup{T}\vdash360_{\pi}^{\circ}540^{\circ}360_{\pi}^{\circ}90^{\circ}$. The reference Rabi frequency for all pulses is $\Omega=\pi/2\tau$ ($\Omega\tau\equiv90^{\circ}$ area in degrees) where the pulse duration reference is $\tau=3/16$~s, the free evolution time is T$=2$~s.}
\label{fig:GHR-interferences}
\end{figure}

\section{COMPOSITE PHASE-SHIFT OPTIMIZATION}

\subsection{Analytical formula}

\indent The Pauli-spin model is now tested in cases when composite pulses are used in Ramsey interferometry.
As a demonstration, a few composite phase-shifts are derived following our recursive algorithm.
The quantization axis is oriented along the z axis as $\widehat{m}=(0,0,1)$ for laser pulsed qubit spectroscopy. Normalized unitary rotation axis parameters can be introduced as $\widehat{n}_{l_{x}}\equiv\frac{\Omega_{l}}{\omega_{l}}\cos\varphi_{l}$, $\widehat{n}_{l_{y}}\equiv\frac{\Omega_{l}}{\omega_{l}}\sin\varphi_{l}$ and $\widehat{n}_{l_{z}}\equiv\frac{\delta_{l}}{\omega_{l}}$ ($l\in \{p,q\}$) respectively related to complex Rabi field frequency in the $x,y$ plane and frequency detuning along the $z$ axis~\cite{Shoemaker:1978}. We therefore define the effective Rabi field as $\omega_{l}=\sqrt{\delta_{l}^{2}+\Omega_{l}^{2}}$.

In selected interrogation schemes reported in Fig.~\ref{fig:protocols}, dispersive error signals of spinor interferences are produced by subtracting two recorded transition probabilities $_{p}^{q}\textup{P}_{gg}$ with additional laser phase-steps $\pm\varphi_{l}$ opposite in sign and applied on required pulses to produce dispersive curves or interferences~\cite{Ramsey:1951,Zanon-Willette:2018}:
\begin{equation}
\begin{split}
\Delta\textup{E}=_{p}^{q}\textup{P}_{gg}(+\varphi_{l})-_{p}^{q}\textup{P}_{gg}(-\varphi_{l}),\label{eq:clock-interference}
\end{split}
\end{equation}
While addressing the $_{p}^{q}C_{gg}$ matrix element with Pauli matrices, the atomic phase-shift expression, for a generalized hyper-Ramsey interference with $\{p,q\}$ composite pulses, can always be decomposed into two contributions:
\begin{equation}
\begin{split}
_{p}^{q}\widetilde{\Phi}_{gg}^{+}=&\arctan\left[\frac{\left(\right)^{p}_{z}
+\left(\right)^{q}_{z}}{1-\left(\right)^{p}_{z}\left(\right)^{q}_{z}}\right]\\
_{p}^{q}\widetilde{\Phi}_{gg}^{-}=&\arctan\left[\frac{\left(\right)^{p}_{y}\left(\right)^{q}_{x}-\left(\right)^{p}_{x}\left(\right)^{q}_{y}}
{\left(\right)^{p}_{x}\left(\right)^{q}_{x}+\left(\right)^{p}_{y}\left(\right)^{q}_{y}}\right]
\end{split}
\label{eq:HR-RG-phase-shift}
\end{equation}
Few elements $\left(\right)^{p,q}_{x,y,z}$ will be given later.
Note that $_{p}^{q}\widetilde{\Phi}_{gg}^{\pm}$ can be recast into a single canonical expression as~\cite{Abramowitz:1968}:
\begin{equation}
\begin{split}
_{p}^{q}\widetilde{\Phi}_{gg}^{+}\mp_{p}^{q}\widetilde{\Phi}_{gg}^{-}=\arctan\left[\frac{\tan_{p}^{q}\widetilde{\Phi}_{gg}^{+}\mp\tan_{p}^{q}\widetilde{\Phi}_{gg}^{-}}{1\pm\tan_{p}^{q}\widetilde{\Phi}_{gg}^{+}\tan_{p}^{q}\widetilde{\Phi}_{gg}^{-}}\right]
\end{split}
\label{eq:recast}
\end{equation}
\indent Various interrogation protocols are now investigated. Two-pulse, three-pulse and five-pulse protocols are shown in the diagram of Fig.~\ref{fig:protocols}; they can be identified by the rotation angle of each pulse, expressed in terms of an integer multiple of $90^{\circ}$.

Using this approach, pulses are classified as Ramsey $[1:1]$ (R, blue), Hyper-Ramsey $[1:2:1]$ (HR3$_\pi$, green) and high-order hyper-Ramsey $[1:4:6:4:1]$ (HR5$_\pi$, red) protocols following the Pascal's triangle for binomial coefficients representation. They can be symmetrically read from left to right or from right to left in the diagram of Fig.~\ref{fig:protocols}.
\begin{figure}[t!!]
\center
\resizebox{8.5cm}{!}{\includegraphics[angle=0]{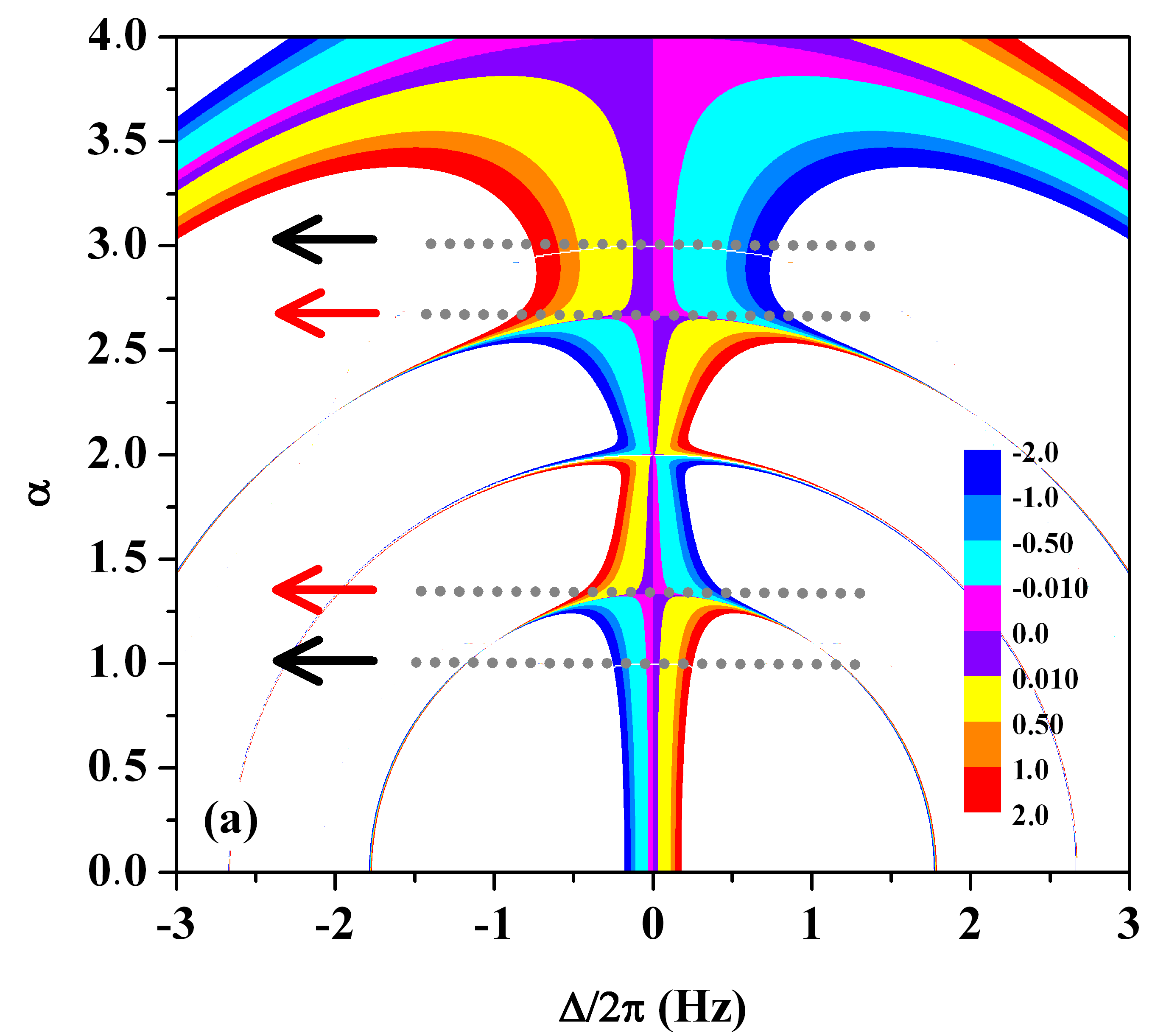}}
\resizebox{8.5cm}{!}{\includegraphics[angle=0]{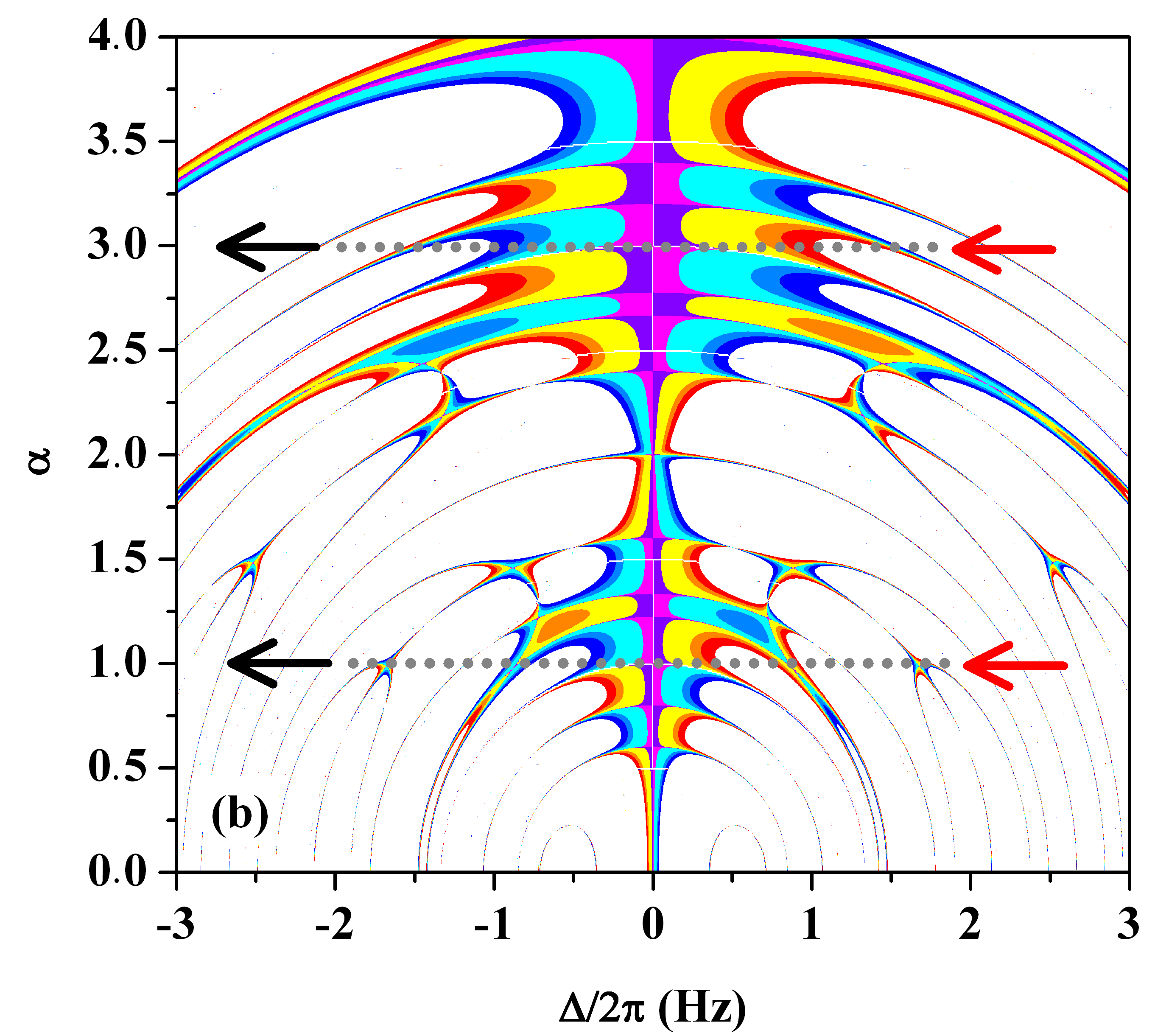}}
\caption{(color online). 2D diagrams of the hyper-Ramsey clock frequency-shift (a) $_{1'}^{2}\widetilde{\Phi}_{gg}/(2\pi\textup{T})$ and (b) $_{1'}^{4}\widetilde{\Phi}_{gg}/(2\pi\textup{T})$ versus uncompensated part of a residual light-shift $\Delta/2\pi$ along the horizontal axis and pulse area $\Omega\tau$ along the vertical axis (see also~\cite{Zanon-Willette:2018}). The only fixed parameter is the free evolution time as $\textup{T}=2$~s. Amplitude of the clock frequency-shift is indicated by a color graded scale from -2~mHz to 2~mHz on the right side. White regions correspond to values out of range. The reference pulse is $\Omega\tau=\alpha\times\pi/2$ where $\alpha$ is the parameter tuned along the vertical axis multiplying all single pulse areas within a composite pulse protocol. Phase-shifts are evaluated modulo $\pm k\pi,~k\in\mathbb{N}$ (see also~\cite{Zanon-Willette:2019}). Arrows are pointing pulse area values for a maximum central fringe amplitude (black) or a high-order sensitivity to the accumulated phase-shift (red).}
\label{fig:2D-map-GHR}
\end{figure}
This classification is used to identify pulse protocols which make the optical clock sensitive only at high-order levels to the probe laser frequency shifts where the degree of sensitivity scales with the number of pulses. This is also remnant to the Taylor expansion of some leading coefficients of pulse parameters from the theoretical analysis of refs~\cite{Yudin:2010,Yudin:2016}.
The free-rotation of the qubit denoted $\delta\textup{T}$ has to be positioned between the first two pulses (or the last two pulses) of each configuration. If not, other protocols are generated with a different interferometric lineshape and sensitivity to the residual light-shift (see for example two additional examples of spinor interferences with four pulses shown in Fig.~\ref{fig:spinor-interferences} from the appendix).

The two-pulse Ramsey (R) protocol ($p=1',q=1$) was proposed in 1949~\cite{Ramsey:1949}. The hyper-Ramsey interrogation scheme ($p=2',q=1$ or $p=1',q=2$) originally presented in 2010, denoted as HR3$_{\pi}$ protocol, is based on a sequence of three laser pulses~\cite{Yudin:2010}. It relies on replacing the first or the second Ramsey pulse by a combination of two pulses (a composite pulse) including an additional laser phase-step of $\pi$. The sequence of five laser pulses ($p=4',q=1$ or $p=1',q=4$) is a new high-order HR5$_{\pi}$ protocol including this time a set of more elaborated composite pulses as $360_{\pi}^{\circ}540^{\circ}360_{\pi}^{\circ}$ replacing the intermediate $180_{\pi}^{\circ}$ pulse. For both cases HR3$_\pi$ and HR5$_\pi$, the interference signal calculated using Eq.~\ref{eq:clock-interference} is shown in Fig.~\ref{fig:GHR-interferences}(a) and (b) versus the clock detuning.
The three-pulse protocol generating hyper-Ramsey interferences (Fig.~\ref{fig:GHR-interferences}(a)) has been successfully applied on the single-ion $^{171}$Yb$^{+}$ octupole clock demonstrating a relative accuracy of $3\times10^{-18}$~\cite{Huntemann:2016}.
The composite phase-shift related to these configurations is denoted $_{p}^{q}\widetilde{\Phi}_{gg}$. To derive the analytical expression of the corresponding clock frequency-shift $_{p}^{q}\widetilde{\Phi}/(2\pi \textup{T})$, the ERG transformation rules are iterated up to three times generating required $\left(\right)^{p,q}_{x,y,z}$ elements with $l=1',2',3';l=1,2,3\in\{4',4\}$.

\subsection{2D map optimization of composite pulse protocols}

\indent  As applications, we consider relevant phase-shifts $_{1'}^{1}\widetilde{\Phi}_{gg}$, $_{1'}^{2}\widetilde{\Phi}_{gg}$ and $_{1'}^{4}\widetilde{\Phi}_{gg}$. Corresponding 2D diagrams reconstructing the clock frequency-shifts $_{1'}^{2}\widetilde{\Phi}_{gg}/(2\pi\textup{T})$ and $_{1'}^{4}\widetilde{\Phi}_{gg}/(2\pi\textup{T})$ versus the residual light-shift and pulse area are shown in Fig.~\ref{fig:2D-map-GHR}(a) and (b). We choose the amplitude of the frequency-shift to be indicated by a color graded scale between -2~mHz and 2~mHz values as in ~\cite{Zanon-Willette:2017}.

A careful investigation of these diagrams allows us to extract some key-parameters optimizing the robustness of the clock frequency-shift associated to HR3$_{\pi}$ and HR5$_{\pi}$ protocols.
We are able to explore wide regions of several multiple values of the laser pulse area where the single pulse area reference is introduced as $\Omega\tau=\alpha\times\pi/2$ ($\equiv90^{\circ}$). The error signal
amplitude is always maximized for odd values and vanishing for even values of this $\alpha$ parameter.
By increasing the pulse area tuning the parameter $\alpha$, the related light-shift correction is increasing quadratically with the Rabi frequency but can still be fully compensated by adjusting
the laser frequency-step~\cite{Yudin:2009}.
\begin{figure}[t!!]
\center
\resizebox{8.5cm}{!}{\includegraphics[angle=0]{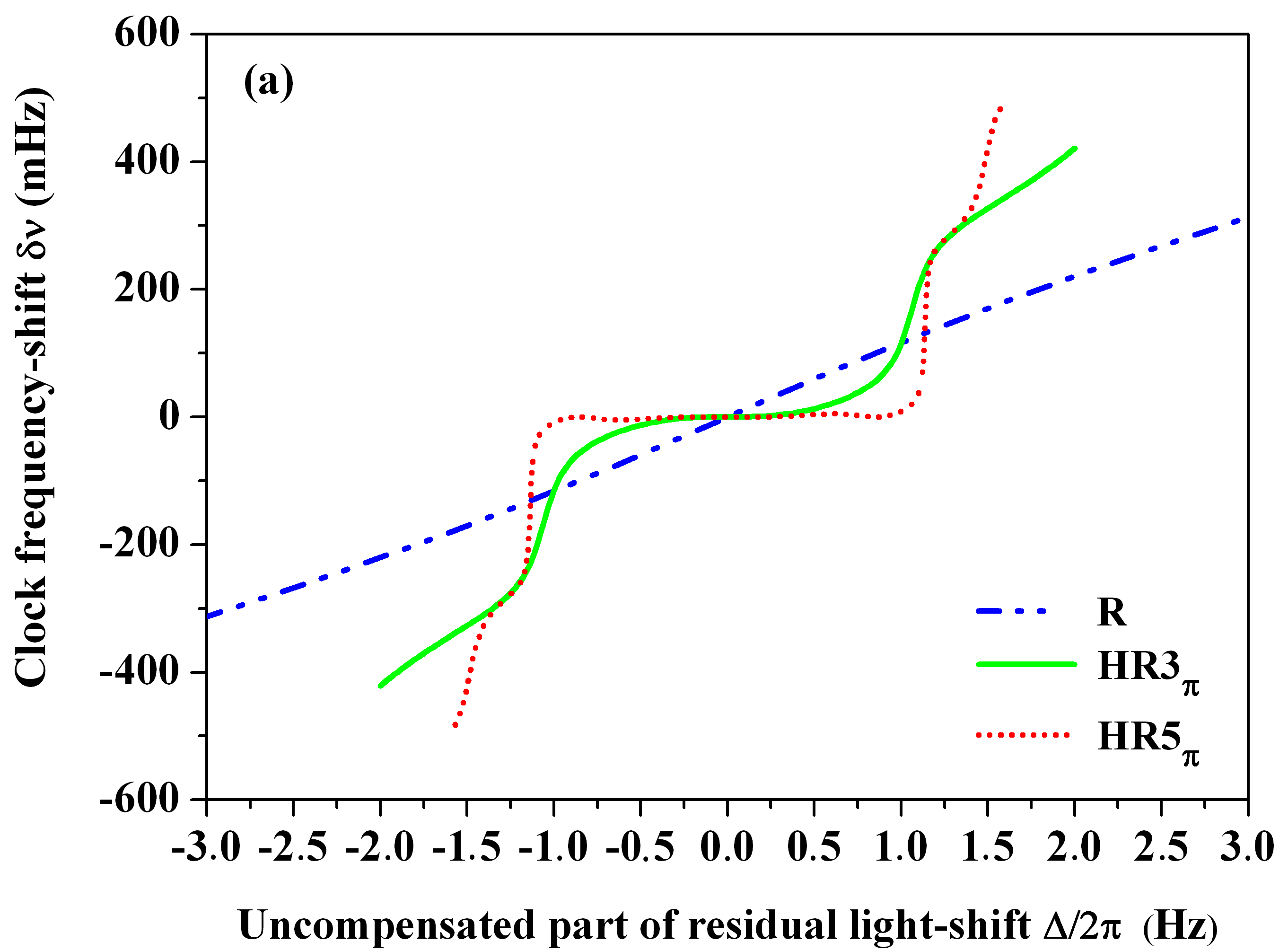}}
\resizebox{8.5cm}{!}{\includegraphics[angle=0]{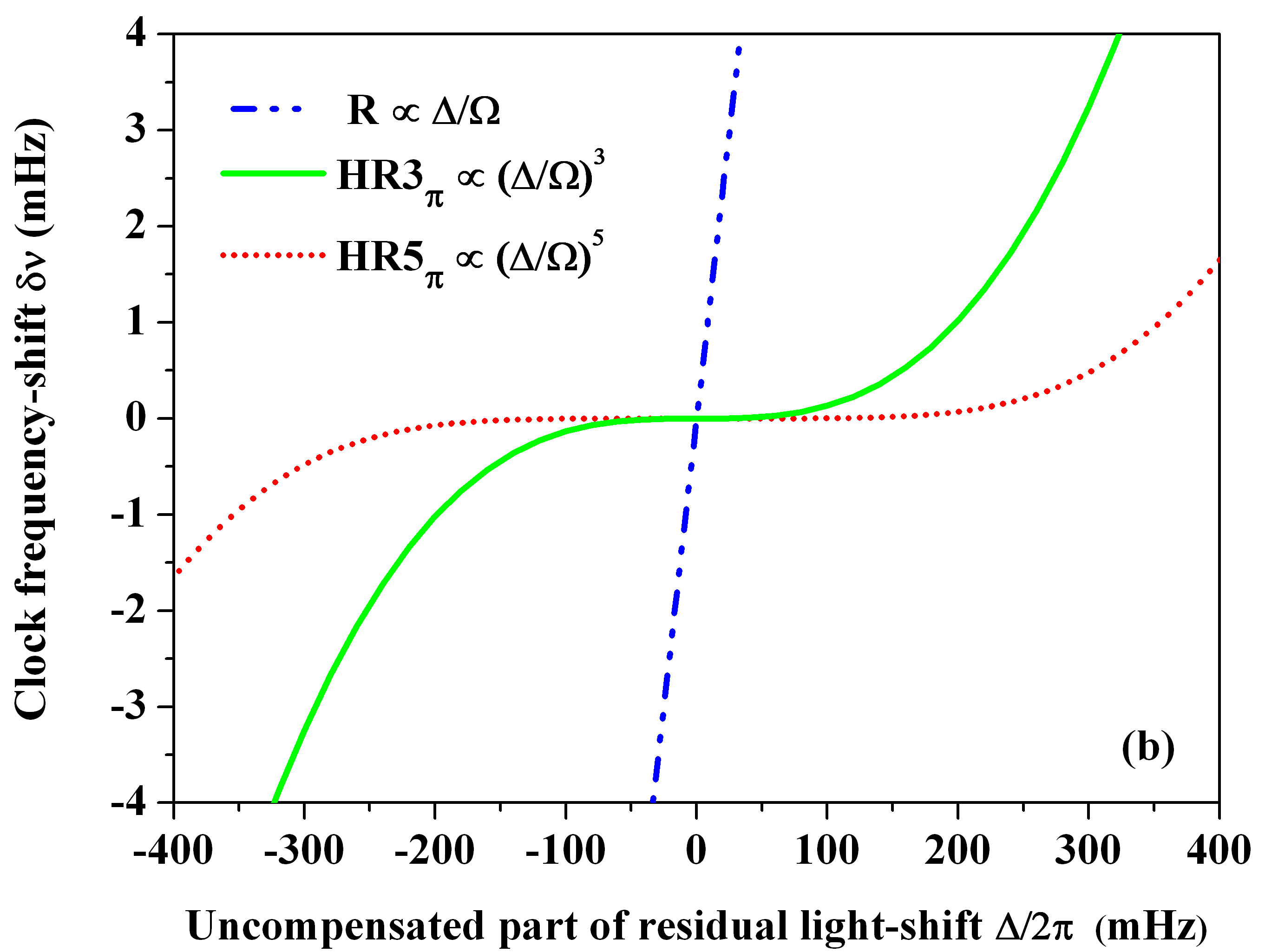}}
\caption{(color online). Central error signal interference frequency-shift $_{p}^{q}\widetilde{\Phi}_{gg}/(2\pi\textup{T})$ versus residual uncompensated part of the light-shift (a) for Ramsey (R) protocol ($_{1'}^{1}\widetilde{\Phi}_{gg}$ with blue dash-dotted line), HR3$_{\pi}$ protocol ($_{1'}^{2}\widetilde{\Phi}_{gg}$ with a continuous green line) and HR5$_{\pi}$ protocol ($_{1'}^{4}\widetilde{\Phi}_{gg}$ with a red short-dotted line). (b) Zoom of clock frequency-shifts emphasizing the linear dependence $\propto\Delta/\Omega$, the cubic dependence $\propto(\Delta/\Omega)^{3}$ and the quintic dependence $ \propto(\Delta/\Omega)^{5}$ versus residual uncompensated part of the light-shift $\Delta/2\pi$. Same laser parameters as in Fig.~\ref{fig:GHR-interferences} with a fixed pulse area parameter $\alpha=1$ for all curves.}
\label{fig:frequency-shifts}
\end{figure}
The clock-frequency-shift compensation can be made more robust by slightly shifting the pulse area parameter from $\alpha=1,3$ (black arrow) to $\alpha=4/3,8/3$ (red arrows) for the HR3$_{\pi}$ protocol (black and red arrows are pointing different pulse area values reported in Fig.~\ref{fig:2D-map-GHR}(a)). These particular pulse areas are strongly upgrading the compensation of the clock frequency-shift up to two or three orders of magnitude but are associated to a relative reduction in the central fringe amplitude by $\sim30\%$. As an example, injecting a residual light-shift of $\Delta/2\pi\sim200$~mHz with a free evolution time of T$=2$~s using the HR3$_{\pi}$ protocol, the composite clock frequency-shift $_{1'}^{2}\widetilde{\Phi}_{gg}/(2\pi\textup{T})$ drops from $1.3$~mHz (for $\alpha=1$) to -$0.043$~mHz (for $\alpha=4/3$) and even -$2.5$~$\mu$Hz (for $\alpha=8/3$) if adjusting carefully the light-shift pre-compensation.
Injecting the same initial residual light-shift with the HR5$_{\pi}$ protocol gives a clock frequency-shift $_{1'}^{4}\widetilde{\Phi}_{gg}/(2\pi\textup{T})$ about $0.11$~mHz (for $\alpha=1$) collapsing to $1.5$~$\mu$Hz (for $\alpha=3$) with no reduction in the signal amplitude.

We note that these specific values of the pulse area are steering the clock frequency-shift to locking-points offering a better compensation of the systematics synchronized with the laser probe intensity fluctuation when a pulse area modification of a few percents as $\Delta\theta/\theta\equiv\Delta\alpha/\alpha<\pm5\%$ is tolerated (identified by violet and pink color graded regions in Fig.~\ref{fig:2D-map-GHR}(a) and (b) where an abrupt flip is observed between positive and negative clock frequency-shift values of similar amplitude).
As a difference to the HR3$_{\pi}$ scheme, the HR5$_{\pi}$ protocol gets exceptional pulse area values where the fringe interference contrast is maximum and the clock frequency-shift exhibits a higher order dependence to the residual light-shift (black and red arrows are pointing to the same pulse area values denoted as exceptional points reported in Fig.~\ref{fig:2D-map-GHR}(b)). Our new HR5$_{\pi}$ protocol is simultaneously optimizing the error signal amplitude and the robustness of quantum interferences against probe-induced uncompensated residual light-shifts.

Clock frequency-shifts of quantum interferences versus the residual light-shift are finally simulated and reported in Fig.~\ref{fig:frequency-shifts}(a) and (b) with a reference pulse fixed to $\Omega\tau=\pi/2$  ($\equiv90^{\circ}$) following the horizontal axis of Fig.~\ref{fig:2D-map-GHR} taking $\alpha=1$). While the Ramsey clock frequency-shift is a linear function of the residual light-shift affecting the quantum states, the cubic sensitivity from a three-pulse scheme turns to collapse to a high-order quintic sensitivity to residual light-shifts under a five-pulse protocol as reported in Fig.~\ref{fig:frequency-shifts}(b).

The required elements needed to calculate $_{1'}^{1}\widetilde{\Phi}_{gg}$ are given by:
\begin{equation}
\begin{split}
\begin{array}{c}
\left(\right)^{1'}_{x}=\widehat{n}_{1'_{x}}\tan\widetilde{\theta}_{1'},\hspace{0.5cm}\left(\right)^{1}_{x}=\widehat{n}_{1_{x}}\tan\widetilde{\theta}_{1}\\
\left(\right)^{1'}_{y}=\widehat{n}_{1'_{y}}\tan\widetilde{\theta}_{1'},\hspace{0.5cm}\left(\right)^{1}_{y}=\widehat{n}_{1_{y}}\tan\widetilde{\theta}_{1}\\
\left(\right)^{1'}_{z}=\widehat{n}_{1'_{z}}\tan\widetilde{\theta}_{1'},\hspace{0.5cm}\left(\right)^{1}_{z}=\widehat{n}_{1_{z}}\tan\widetilde{\theta}_{1}
\end{array}
\end{split}
\label{eq:R-transformation}
\end{equation}
By applying Eq.~\ref{eq:recast}, the original Ramsey clock frequency-shift reduces to:
\begin{equation}
\begin{split}
_{1'}^{1}\widetilde{\Phi}_{gg}&=\varphi_{1}-\varphi_{1'}+\frac{\delta_{1'}}{\omega_{1'}}\tan\widetilde{\theta}_{1'}+\frac{\delta_{1}}{\omega_{1}}\tan\widetilde{\theta}_{1}\\
&=\varphi_{1}-\varphi_{1'}+\phi_{1'}+\phi_{1}
\label{eq:phase-shift-R}
\end{split}
\end{equation}
in accordance with~\cite{Zanon-Willette:2015,Zanon-Willette:2019,Zanon-Willette:2016}.

The required elements needed to calculate $_{1'}^{2}\widetilde{\Phi}_{gg}$ or $_{2'}^{1}\widetilde{\Phi}_{gg}$ are:
\begin{equation}
\begin{split}
\begin{array}{c}
\left(\right)^{2'}_{x}=\frac{\widehat{n}_{1'_{x}}\tan\widetilde{\theta}_{1'}
+\widehat{n}_{2'_{x}}\tan\widetilde{\theta}_{2'}+(\widehat{n}_{1'_{z}}\widehat{n}_{2'_{y}}-\widehat{n}_{1'_{y}}\widehat{n}_{2'_{z}})\tan\widetilde{\theta}_{1'}\tan\widetilde{\theta}_{2'}}
{1-\left(n_{1'_{x}}n_{2'_{x}}+n_{1'_{y}}n_{2'_{y}}+n_{1'_{z}}n_{2'_{z}}\right)\tan\widetilde{\theta}_{1'}\tan\widetilde{\theta}_{2'}}\\
\left(\right)^{2'}_{y}=\frac{\widehat{n}_{1'_{y}}\tan\widetilde{\theta}_{1'}
+\widehat{n}_{2'_{y}}\tan\widetilde{\theta}_{2'}+(\widehat{n}_{1'_{x}}\widehat{n}_{2'_{z}}-\widehat{n}_{1'_{z}}\widehat{n}_{2'_{x}})\tan\widetilde{\theta}_{1'}\tan\widetilde{\theta}_{2'}}
{1-\left(n_{1'_{x}}n_{2'_{x}}+n_{1'_{y}}n_{2'_{y}}+n_{1'_{z}}n_{2'_{z}}\right)\tan\widetilde{\theta}_{1'}\tan\widetilde{\theta}_{2'}}\\
\left(\right)^{2'}_{z}=\frac{\widehat{n}_{1'_{z}}\tan\widetilde{\theta}_{1'}
+\widehat{n}_{2'_{z}}\tan\widetilde{\theta}_{2'}+(\widehat{n}_{1'_{y}}\widehat{n}_{2'_{x}}-\widehat{n}_{1'_{x}}\widehat{n}_{2'_{y}})\tan\widetilde{\theta}_{1'}\tan\widetilde{\theta}_{2'}}
{1-\left(n_{1'_{x}}n_{2'_{x}}+n_{1'_{y}}n_{2'_{y}}+n_{1'_{z}}n_{2'_{z}}\right)\tan\widetilde{\theta}_{1'}\tan\widetilde{\theta}_{2'}}
\end{array}
\end{split}
\label{eq:GHR-2pulses-1}
\end{equation}
and
\begin{equation}
\begin{split}
\begin{array}{c}
\left(\right)^{2}_{x}=\frac{\widehat{n}_{1_{x}}\tan\widetilde{\theta}_{1}
+\widehat{n}_{2_{x}}\tan\widetilde{\theta}_{2}-(\widehat{n}_{1_{z}}\widehat{n}_{2_{y}}-\widehat{n}_{1_{y}}\widehat{n}_{2_{z}})\tan\widetilde{\theta}_{1}\tan\widetilde{\theta}_{2}}
{1-\left(n_{1_{x}}n_{2_{x}}+n_{1_{y}}n_{2_{y}}+n_{1_{z}}n_{2_{z}}\right)\tan\widetilde{\theta}_{1}\tan\widetilde{\theta}_{2}}\\
\left(\right)^{2}_{y}=\frac{\widehat{n}_{1_{y}}\tan\widetilde{\theta}_{1}
+\widehat{n}_{2_{y}}\tan\widetilde{\theta}_{2}-(\widehat{n}_{1_{x}}\widehat{n}_{2_{z}}-\widehat{n}_{1_{z}}\widehat{n}_{2_{x}})\tan\widetilde{\theta}_{1}\tan\widetilde{\theta}_{2}}
{1-\left(n_{1_{x}}n_{2_{x}}+n_{1_{y}}n_{2_{y}}+n_{1_{z}}n_{2_{z}}\right)\tan\widetilde{\theta}_{1}\tan\widetilde{\theta}_{2}}\\
\left(\right)^{2}_{z}=\frac{\widehat{n}_{1_{z}}\tan\widetilde{\theta}_{1}
+\widehat{n}_{2_{z}}\tan\widetilde{\theta}_{2}-(\widehat{n}_{1_{y}}\widehat{n}_{2_{x}}-\widehat{n}_{1_{x}}\widehat{n}_{2_{y}})\tan\widetilde{\theta}_{1}\tan\widetilde{\theta}_{2}}
{1-\left(n_{1_{x}}n_{2_{x}}+n_{1_{y}}n_{2_{y}}+n_{1_{z}}n_{2_{z}}\right)\tan\widetilde{\theta}_{1}\tan\widetilde{\theta}_{2}}
\end{array}
\end{split}
\label{eq:GHR-2pulses-2}
\end{equation}
By fixing $\theta_{2'}\equiv0$ into Eq.~\ref{eq:GHR-2pulses-1} while inserting $\widehat{n}_{1_{y}}=\widehat{n}_{2_{y}}\equiv0$ in Eq.~\ref{eq:GHR-2pulses-2}, the hyper-Ramsey clock frequency-shift becomes identical to~\cite{Zanon-Willette:2015}.

Required elements needed to calculate $_{1'}^{4}\widetilde{\Phi}_{gg}$ or $_{4'}^{1}\widetilde{\Phi}_{gg}$ are rapidly increasing in size and are not given here. They can be derived applying two-times the ERG transformation rules on Eq.~\ref{eq:GHR-2pulses-1} and Eq.~\ref{eq:GHR-2pulses-2} (see the appendix section for analytics). This five-pulse protocol has been also derived with the other recursive algorithm~\cite{Note} following~\cite{Zanon-Willette:2022} confirming the accuracy of the plots reported in Fig.~\ref{fig:frequency-shifts}(a) and (b).

\section{Conclusion}

\indent A SU(2) formulation of hyper-Ramsey interferences with composite phase-shifts has been presented. Hyper-clock interrogation protocols and their interferometric dependence to light-shift have been classified by analogy with a Pascal's triangle representation of doublet, triplet and quintet splitting patterns from spin-spin interaction in proton NMR multiplet spectroscopy~\cite{Valiulin:2019}. Such a representation may ease the search for new and more efficient interrogation protocols of ultra-narrow optical clock transitions while offering a framework to derive analytically the phase-shifts associated to arbitrary laser pulses, showing the remarkable intrinsic robustness introduced in quantum metrology by hyper qubit-clocks.
In the present work, a five-pulse protocol is discovered to be a high-order version of the hyper-Ramsey three-pulse scheme demonstrating a quintic sensitivity to residual probe-induced light-shifts with a maximum signal amplitude.

The Pauli-spin model, complementary to analytical tools introduced in~\cite{Zanon-Willette:2022} describing hyper-Ramsey-Bordé matter-wave interferometry, uses another recursive algorithm connected to rotation composition rules of unit-quaternions (or versors) algebra in a four dimensional space~\cite{Altmann:1986}.
Natural extension to SU(3) composite phase-shifts via three-level state interferences (hyper qutrit-clock) may be also explored~\cite{Hioe:1981} using a compact representation of Gell-Mann spin matrices~\cite{Gell-Mann:1962,Curtright:2015,Weigert:1997}.
Composite phase-shifts would certainly be an advantage to qudit multiple rotations exposed to detrimental ac Stark-shifts for robust quantum computation~\cite{Ringbauer:2021}.
The next generation of quantum clocks will irrevocably bring a relative level of accuracy below 10$^{-18}$ through very long coherence times~\cite{Brewer:2019,Ye:2008,Hutson:2019}, probably supported by robustness against noise with programmable quantum circuit technologies~\cite{Kaubruegger:2019,Kaubruegger:2021}, quantum non demolition measurements~\cite{Kohlhaas:2015,Bowden:2020} and state entanglement~\cite{Pedrozo:2020}.
At this future level of accuracy, hyper-clocks with an optimal control of composite phase-shifts should reduce the influence of laser-probe-intensity fluctuations~\cite{Beloy:2018} while offering an additional toolbox for a fine tuning control of the optical clock frequency in trapped multi-ion clocks~\cite{Schulte:2016} and in optical lattice clocks~\cite{Ushijima:2018}.

This work in parallel with~\cite{Zanon-Willette:2022} should serve as quantum engineering methods to explore cooperative composite pulse protocols~\cite{Zanon-Willette:2017,Braun:2014} dedicated to robust control algorithms upgrading performances of optical frequency standards~\cite{Ludlow:2015}, quantum computation with qubits and qudits immune to light-shift~\cite{Ringbauer:2021,Wang:2020}, robust quantum sensing~\cite{Degen:2017} and pushing further high-precision laser spectroscopy with cold molecules~\cite{Kondov:2019} and cold anti-matter~\cite{Baker:2021}.

\section{Acknowledgment}

T.Z.W. is deeply grateful to Dr J.-P. Karr, Dr E. de Clercq, Pr M. Cahay, the Wilkowski lab teams with Sr(I) and Sr(II) projects for discussion, comments and criticism.
V.I.Yudin was supported by the Russian Foundation for Basic Research (Grant Nos. 20-02-00505 and 19-32-90181) and Foundation for the Advancement of Theoretical Physics and Mathematics "BASIS".
A.V. Taichenachev acknowledges financial support from Russian Science Foundation through the grant 20-12-00081.
T.Z.W. acknowledges Sorbonne Université and MajuLab for supporting a twelve months visiting research associate professorship at center for quantum technologies (CQT) in Singapore.

\section*{Appendix}

\subsection*{A: building-block for $\tan_{p}^{q}\widetilde{\Phi}_{uu'}^{\pm}$}

\indent In this section, the decomposition of Eq.~\ref{eq:complex-phase-shift} from the main text is explicitly provided with cartesian axis coordinates $\widehat{n}_{p_{x,y,z}}$, $\widehat{n}_{q_{x,y,z}}$ and  $\widehat{m}_{x,y,z}$.
The interferometric composite phase-shift numerator and denominator can be explicitly developed using the Pauli matrices.
For the diagonal phase-shift, we obtain the numerator components to build $\left\{\tan_{p}^{q}\widetilde{\Phi}_{gg}^{\pm}\right\}_{N}$:
\begin{equation}
\begin{split}
\widehat{n}_{p}\cdot\left(\overrightarrow{\sigma}\pm\widehat{m}\sigma_{0}\right)=&\widehat{n}_{p_{z}}\pm(\widehat{m}_{x}\widehat{n}_{p_{x}}+\widehat{m}_{y}\widehat{n}_{p_{y}}+\widehat{m}_{z}\widehat{n}_{p_{z}})\\
\widehat{n}_{q}\cdot\left(\overrightarrow{\sigma}\pm\widehat{m}\sigma_{0}\right)=&\widehat{n}_{q_{z}}\pm(\widehat{m}_{x}\widehat{n}_{q_{x}}+\widehat{m}_{y}\widehat{n}_{q_{y}}+\widehat{m}_{z}\widehat{n}_{q_{z}})\\
\left(\widehat{n}_{p}\times\widehat{n}_{q}\right)\cdot\left(\overrightarrow{\sigma}\mp\widehat{m}\sigma_{0}\right)=&\widehat{n}_{p_{x}}\widehat{n}_{q_{y}}-\widehat{n}_{q_{x}}\widehat{n}_{p_{y}}\\
&\mp\widehat{m}_{x}(\widehat{n}_{q_{z}}\widehat{n}_{p_{y}}-\widehat{n}_{q_{y}}\widehat{n}_{p_{z}})\\
&\mp\widehat{m}_{y}(\widehat{n}_{q_{x}}\widehat{n}_{p_{z}}-\widehat{n}_{q_{z}}\widehat{n}_{p_{x}})\\
&\mp\widehat{m}_{z}(\widehat{n}_{p_{x}}\widehat{n}_{q_{y}}-\widehat{n}_{q_{x}}\widehat{n}_{p_{y}})
\end{split}
\label{eq:D1}
\end{equation}
and the denominator components to build $\left\{\tan_{p}^{q}\widetilde{\Phi}_{gg}^{\pm}\right\}_{D}$:
\begin{equation}
\begin{split}
\sigma_{0}\pm\widehat{m}\overrightarrow{\sigma}=&1\pm\widehat{m}_{z}\\
\left[\widehat{n}_{p}\times\widehat{m}\right]\cdot\overrightarrow{\sigma}=&\widehat{m}_{y}\widehat{n}_{p_{x}}-\widehat{m}_{x}\widehat{n}_{p_{y}}\\
\left[\widehat{n}_{q}\times\widehat{m}\right]\cdot\overrightarrow{\sigma}=&\widehat{m}_{y}\widehat{n}_{q_{x}}-\widehat{m}_{x}\widehat{n}_{q_{y}}\\
\left(\widehat{n}_{p}\cdot\widehat{n}_{q}\right)_{\widehat{m},\overrightarrow{\sigma}}=&\widehat{n}_{p_{x}}\widehat{n}_{q_{x}}+\widehat{n}_{p_{y}}\widehat{n}_{q_{y}}+\widehat{n}_{p_{z}}\widehat{n}_{q_{z}}\\
&\mp\widehat{m}_{z}\left(\widehat{n}_{p_{x}}\widehat{n}_{q_{x}}+\widehat{n}_{p_{y}}\widehat{n}_{q_{y}}+\widehat{n}_{p_{z}}\widehat{n}_{q_{z}}\right)\\
&\pm\widehat{n}_{q_{z}}\left(\widehat{m}_{x}\widehat{n}_{p_{x}}+\widehat{m}_{y}\widehat{n}_{p_{y}}+\widehat{m}_{z}\widehat{n}_{p_{z}}\right)\\
&\pm\widehat{n}_{p_{z}}\left(\widehat{m}_{x}\widehat{n}_{q_{x}}+\widehat{m}_{y}\widehat{n}_{q_{y}}+\widehat{m}_{z}\widehat{n}_{q_{z}}\right)
\end{split}
\label{eq:D2}
\end{equation}
For the off-diagonal complex phase-shift, we obtain the numerator components for $\left\{\tan_{p}^{q}\widetilde{\Phi}_{eg}^{\pm}\right\}_{N}$:
\begin{equation}
\begin{split}
\widehat{n}_{p}\cdot\left(\overrightarrow{\sigma}\pm\widehat{m}\sigma_{0}\right)=&\widehat{n}_{p_{x}}+i~\widehat{n}_{p_{y}}\\
\widehat{n}_{q}\cdot\left(\overrightarrow{\sigma}\pm\widehat{m}\sigma_{0}\right)=&\widehat{n}_{q_{x}}+i~\widehat{n}_{q_{y}}\\
\left(\widehat{n}_{p}\times\widehat{n}_{q}\right)\cdot\left(\overrightarrow{\sigma}\mp\widehat{m}\sigma_{0}\right)=&\left(\widehat{n}_{q_{y}}\widehat{n}_{p_{z}}-\widehat{n}_{q_{z}}\widehat{n}_{p_{y}}\right)\\
&+i~\left(\widehat{n}_{q_{z}}\widehat{n}_{p_{x}}-\widehat{n}_{q_{x}}\widehat{n}_{p_{z}}\right)
\end{split}
\label{eq:OD1}
\end{equation}
and the denominator components for $\left\{\tan_{p}^{q}\widetilde{\Phi}_{eg}^{\pm}\right\}_{D}$:
\begin{equation}
\begin{split}
\sigma_{0}\pm\widehat{m}\overrightarrow{\sigma}=&\pm(\widehat{m}_{x}+i~\widehat{m}_{y})\\
\left[\widehat{n}_{p}\times\widehat{m}\right]\cdot\overrightarrow{\sigma}=&(\widehat{m}_{z}\widehat{n}_{p_{y}}-\widehat{m}_{y}\widehat{n}_{p_{z}})+i~\left(\widehat{m}_{x}\widehat{n}_{p_{z}}-\widehat{m}_{z}\widehat{n}_{p_{x}}\right)\\
\left[\widehat{n}_{q}\times\widehat{m}\right]\cdot\overrightarrow{\sigma}=&(\widehat{m}_{z}\widehat{n}_{q_{y}}-\widehat{m}_{y}\widehat{n}_{q_{z}})+i~\left(\widehat{m}_{x}\widehat{n}_{q_{z}}-\widehat{m}_{z}\widehat{n}_{q_{x}}\right)\\
\left(\widehat{n}_{p}\cdot\widehat{n}_{q}\right)_{\widehat{m},\overrightarrow{\sigma}}=&\mp\left(\widehat{m}_{x}+i~\widehat{m}_{y}\right)\left(\widehat{n}_{p_{x}}\widehat{n}_{q_{x}}+\widehat{n}_{p_{y}}\widehat{n}_{q_{y}}+\widehat{n}_{p_{z}}\widehat{n}_{q_{z}}\right)\\
&\pm\left(\widehat{n}_{q_{x}}+i~\widehat{n}_{q_{y}}\right)\left(\widehat{m}_{x}\widehat{n}_{p_{x}}+\widehat{m}_{y}\widehat{n}_{p_{y}}+\widehat{m}_{z}\widehat{n}_{p_{z}}\right)\\
&\pm\left(\widehat{n}_{p_{x}}+i~\widehat{n}_{p_{y}}\right)\left(\widehat{m}_{x}\widehat{n}_{q_{x}}+\widehat{m}_{y}\widehat{n}_{q_{y}}+\widehat{m}_{z}\widehat{n}_{q_{z}}\right)
\end{split}
\label{eq:OD2}
\end{equation}
where $N,D$ stands for the numerator and the denominator of the quantity $\tan_{p}^{q}\widetilde{\Phi}_{uu'}^{\pm}$ and all elements have to be associated to $\tan\widetilde{\theta}_{p}$ and $\tan\widetilde{\theta}_{q}$.
Now, we proceed by fixing the orientation axis $\widehat{m}=(0,0,1)$ as in the main text. We explicitly derive the diagonal phase-shift expressions $_{p}^{q}\widetilde{\Phi}_{gg}^{\pm}$ with the help of Eq.~\ref{eq:D1} and Eq.~\ref{eq:D2}:
\begin{equation}
\begin{split}
\tan_{p}^{q}\widetilde{\Phi}_{gg}^{+}&=\frac{\widehat{n}_{p_{z}}\tan\widetilde{\theta}_{p}+\widehat{n}_{q_{z}}\tan\widetilde{\theta}_{q}}{1-\widehat{n}_{p_{z}}\widehat{n}_{q_{z}}\tan\widetilde{\theta}_{p}\tan\widetilde{\theta}_{q}}\\
\tan_{p}^{q}\widetilde{\Phi}_{gg}^{-}&=\frac{\widehat{n}_{p_{y}}\widehat{n}_{q_{x}}-\widehat{n}_{p_{x}}\widehat{n}_{q_{y}}}{\widehat{n}_{p_{x}}\widehat{n}_{q_{x}}+\widehat{n}_{p_{y}}\widehat{n}_{q_{y}}}
\end{split}
\end{equation}
Using normalized parameters from the main text $\widehat{n}_{l_{x}}\equiv\frac{\Omega_{l}}{\omega_{l}}\cos\varphi_{l}$, $\widehat{n}_{l_{y}}\equiv\frac{\Omega_{l}}{\omega_{l}}\sin\varphi_{l}$ and $\widehat{n}_{l_{z}}\equiv\frac{\delta_{l}}{\omega_{l}}$ with ($l=p,q$), we obtain, with Eq.~\ref{eq:recast}, the overall Ramsey phase-shift $_{p}^{q}\widetilde{\Phi}_{gg}$:
\begin{equation}
\begin{split}
_{p}^{q}\widetilde{\Phi}_{gg}&=\varphi_{q}-\varphi_{p}+\phi_{p}+\phi_{q}\\
&=\varphi_{1}-\varphi_{1'}+\phi_{1'}+\phi_{1}
\end{split}
\end{equation}
where we use $\phi_{l}=\frac{\delta_{l}}{\omega_{l}}\tan\widetilde{\theta}_{l}$. Indeed, we have recovered the Ramsey phase-shift by fixing $p=1'$ and $q=1$ as two single pulses.

\subsection*{B: $\left(\right)^{4'}_{x,y,z}$ and $\left(\right)^{4}_{x,y,z}$ elements for $_{1'}^{4}\widetilde{\Phi}_{gg}$ and $_{4'}^{1}\widetilde{\Phi}_{gg}$}

\indent The ERG transformation through Eq.(15) is applied twice on numerator and denominator elements from Eq.~\ref{eq:GHR-2pulses-1} and Eq.~\ref{eq:GHR-2pulses-2} with $p=4',q=4$ pulses.

The transformation gives for the set of $p=4'$ composite pulses:
\begin{eqnarray}
\left\{
\begin{split}
\widehat{n}_{2'}\tan\widetilde{\theta}_{2'}&\mapsto\frac{_{2'}^{3'}\widehat{N}_{+}-~_{2'}^{3'}\widehat{N}_{\times}}{1-_{2'}^{3'}\widehat{N}^{0}_{\bullet}}\\
\widehat{n}_{3'}\tan\widetilde{\theta}_{3'}&\mapsto\frac{_{3'}^{4'}\widehat{N}_{+}-~_{3'}^{4'}\widehat{N}_{\times}}{1-_{3'}^{4'}\widehat{N}^{0}_{\bullet}}\\
\end{split}
\right.
\label{eq:}
\end{eqnarray}
The transformation gives for the set of $q=4$ composite pulses:
\begin{eqnarray}
\left\{
\begin{split}
\widehat{n}_{2}\tan\widetilde{\theta}_{2}\mapsto\frac{_{2}^{3}\widehat{N}_{+}+~_{2}^{3}\widehat{N}_{\times}}{1-_{2}^{3}\widehat{N}^{0}_{\bullet}}\\
\widehat{n}_{3}\tan\widetilde{\theta}_{3}\mapsto\frac{_{3}^{4}\widehat{N}_{+}+~_{3}^{4}\widehat{N}_{\times}}{1-_{3}^{4}\widehat{N}^{0}_{\bullet}}\\
\end{split}
\right.
\label{eq:}
\end{eqnarray}
New expressions for components are thus:
\begin{widetext}
\begin{equation}
\begin{split}
\widehat{n}_{2'_{x}}\tan\widetilde{\theta}_{2'}&\mapsto
\frac{\widehat{n}_{2'_{x}}\tan\widetilde{\theta}_{2'}
+\widehat{n}_{3'_{x}}\tan\widetilde{\theta}_{3'}+\left(\widehat{n}_{2'_{z}}\cdot\widehat{n}_{3'_{y}}\tan\widetilde{\theta}_{3'}-\widehat{n}_{2'_{y}}\cdot\widehat{n}_{3'_{z}}\tan\widetilde{\theta}_{3'}\right)\tan\widetilde{\theta}_{2'}}
{1-\left(\widehat{n}_{2'_{x}}\cdot\widehat{n}_{3'_{x}}\tan\widetilde{\theta}_{3'}+\widehat{n}_{2'_{y}}\cdot\widehat{n}_{3'_{y}}\tan\widetilde{\theta}_{3'}+\widehat{n}_{2'_{z}}\cdot\widehat{n}_{3'_{z}}\tan\widetilde{\theta}_{3'}\right)\tan\widetilde{\theta}_{2'}}\\
\widehat{n}_{2'_{y}}\tan\widetilde{\theta}_{2'}&\mapsto
\frac{\widehat{n}_{2'_{y}}\tan\widetilde{\theta}_{2'}
+\widehat{n}_{3'_{y}}\tan\widetilde{\theta}_{3'}+\left(\widehat{n}_{2'_{x}}\cdot\widehat{n}_{3'_{z}}\tan\widetilde{\theta}_{3'}-\widehat{n}_{2'_{z}}\cdot\widehat{n}_{3'_{x}}\tan\widetilde{\theta}_{3'}\right)\tan\widetilde{\theta}_{2'}}
{1-\left(\widehat{n}_{2'_{x}}\cdot\widehat{n}_{3'_{x}}\tan\widetilde{\theta}_{3'}+\widehat{n}_{2'_{y}}\cdot\widehat{n}_{3'_{y}}\tan\widetilde{\theta}_{3'}+\widehat{n}_{2'_{z}}\cdot\widehat{n}_{3'_{z}}\tan\widetilde{\theta}_{3'}\right)\tan\widetilde{\theta}_{2'}}\\
\widehat{n}_{2'_{z}}\tan\widetilde{\theta}_{2'}&\mapsto
\frac{\widehat{n}_{2'_{z}}\tan\widetilde{\theta}_{2'}
+\widehat{n}_{3'_{z}}\tan\widetilde{\theta}_{3'}+\left(\widehat{n}_{2'_{y}}\cdot\widehat{n}_{3'_{x}}\tan\widetilde{\theta}_{3'}-\widehat{n}_{2'_{x}}\cdot\widehat{n}_{3'_{y}}\tan\widetilde{\theta}_{3'}\right)\tan\widetilde{\theta}_{2'}}
{1-\left(\widehat{n}'_{2_{x}}\cdot\widehat{n}_{3'_{x}}\tan\widetilde{\theta}_{3'}+\widehat{n}_{2'_{y}}\cdot\widehat{n}_{3'_{y}}\tan\widetilde{\theta}_{3'}+\widehat{n}_{2'_{z}}\cdot\widehat{n}_{3'_{z}}\tan\widetilde{\theta}_{3'}\right)\tan\widetilde{\theta}_{2'}}
\end{split}
\label{eq:ERG-p-1}
\end{equation}
where $\widehat{n}_{3'_{x,y,z}}\tan\widetilde{\theta}_{3'}$ axial components are replaced by:
\begin{equation}
\begin{split}
\widehat{n}_{3'_{x}}\tan\widetilde{\theta}_{3'}&\mapsto\frac{\widehat{n}_{3'_{x}}\tan\widetilde{\theta}_{3'}
+\widehat{n}_{4'_{x}}\tan\widetilde{\theta}_{4'}+\left(\widehat{n}_{3'_{z}}\cdot\widehat{n}_{4'_{y}}\tan\widetilde{\theta}_{4'}-\widehat{n}_{3'_{y}}\cdot\widehat{n}_{4'_{z}}\tan\widetilde{\theta}_{4'}\right)\tan\widetilde{\theta}_{3'}}
{1-\left(\widehat{n}_{3'_{x}}\cdot\widehat{n}_{4'_{x}}\tan\widetilde{\theta}_{4'}+\widehat{n}_{3'_{y}}\cdot\widehat{n}_{4'_{y}}\tan\widetilde{\theta}_{4'}+\widehat{n}_{3'_{z}}\cdot\widehat{n}_{4'_{z}}\tan\widetilde{\theta}_{4'}\right)\tan\widetilde{\theta}_{3'}}\\
\widehat{n}_{3'_{y}}\tan\widetilde{\theta}_{3'}&\mapsto\frac{\widehat{n}_{3'_{y}}\tan\widetilde{\theta}_{3'}
+\widehat{n}_{4'_{y}}\tan\widetilde{\theta}_{4'}+\left(\widehat{n}_{3'_{x}}\cdot\widehat{n}_{4'_{z}}\tan\widetilde{\theta}_{4'}-\widehat{n}_{3'_{z}}\cdot\widehat{n}_{4'_{x}}\tan\widetilde{\theta}_{4'}\right)\tan\widetilde{\theta}_{3'}}
{1-\left(\widehat{n}_{3'_{x}}\cdot\widehat{n}_{4'_{x}}\tan\widetilde{\theta}_{4'}+\widehat{n}_{3'_{y}}\cdot\widehat{n}_{4'_{y}}\tan\widetilde{\theta}_{4'}+\widehat{n}_{3'_{z}}\cdot\widehat{n}_{4'_{z}}\tan\widetilde{\theta}_{4'}\right)\tan\widetilde{\theta}_{3'}}\\
\widehat{n}_{3'_{z}}\tan\widetilde{\theta}_{3'}&\mapsto\frac{\widehat{n}_{3'_{z}}\tan\widetilde{\theta}_{3'}
+\widehat{n}_{4'_{z}}\tan\widetilde{\theta}_{4'}+\left(\widehat{n}_{3'_{y}}\cdot\widehat{n}_{4'_{x}}\tan\widetilde{\theta}_{4'}-\widehat{n}_{3'_{x}}\cdot\widehat{n}_{4'_{y}}\tan\widetilde{\theta}_{4'}\right)\tan\widetilde{\theta}_{3'}}
{1-\left(\widehat{n}_{3'_{x}}\cdot\widehat{n}_{4'_{x}}\tan\widetilde{\theta}_{4'}+\widehat{n}_{3'_{y}}\cdot\widehat{n}_{4'_{y}}\tan\widetilde{\theta}_{4'}+\widehat{n}_{3'_{z}}\cdot\widehat{n}_{4'_{z}}\tan\widetilde{\theta}_{4'}\right)\tan\widetilde{\theta}_{3'}}
\end{split}
\label{eq:ERG-p-2}
\end{equation}
\end{widetext}
and
\begin{widetext}
\begin{equation}
\begin{split}
\widehat{n}_{2_{x}}\tan\widetilde{\theta}_{2}&\mapsto
\frac{\widehat{n}_{2_{x}}\tan\widetilde{\theta}_{2}
+\widehat{n}_{3_{x}}\tan\widetilde{\theta}_{3}-\left(\widehat{n}_{2_{z}}\cdot\widehat{n}_{3_{y}}\tan\widetilde{\theta}_{3}-\widehat{n}_{2_{y}}\cdot\widehat{n}_{3_{z}}\tan\widetilde{\theta}_{3}\right)\tan\widetilde{\theta}_{2}}
{1-\left(\widehat{n}_{2_{x}}\cdot\widehat{n}_{3_{x}}\tan\widetilde{\theta}_{3}+\widehat{n}_{2_{y}}\cdot\widehat{n}_{3_{y}}\tan\widetilde{\theta}_{3}+\widehat{n}_{2_{z}}\cdot\widehat{n}_{3_{z}}\tan\widetilde{\theta}_{3}\right)\tan\widetilde{\theta}_{2}}\\
\widehat{n}_{2_{y}}\tan\widetilde{\theta}_{2}&\mapsto
\frac{\widehat{n}_{2_{y}}\tan\widetilde{\theta}_{2}
+\widehat{n}_{3_{y}}\tan\widetilde{\theta}_{3}-\left(\widehat{n}_{2_{x}}\cdot\widehat{n}_{3_{z}}\tan\widetilde{\theta}_{3}-\widehat{n}_{2_{z}}\cdot\widehat{n}_{3_{x}}\tan\widetilde{\theta}_{3}\right)\tan\widetilde{\theta}_{2}}
{1-\left(\widehat{n}_{2_{x}}\cdot\widehat{n}_{3_{x}}\tan\widetilde{\theta}_{3}+\widehat{n}_{2_{y}}\cdot\widehat{n}_{3_{y}}\tan\widetilde{\theta}_{3}+\widehat{n}_{2_{z}}\cdot\widehat{n}_{3_{z}}\tan\widetilde{\theta}_{3}\right)\tan\widetilde{\theta}_{2}}\\
\widehat{n}_{2_{z}}\tan\widetilde{\theta}_{2}&\mapsto
\frac{\widehat{n}_{2_{z}}\tan\widetilde{\theta}_{2}
+\widehat{n}_{3_{z}}\tan\widetilde{\theta}_{3}-\left(\widehat{n}_{2_{y}}\cdot\widehat{n}_{3_{x}}\tan\widetilde{\theta}_{3}-\widehat{n}_{2_{x}}\cdot\widehat{n}_{3_{y}}\tan\widetilde{\theta}_{3}\right)\tan\widetilde{\theta}_{2}}
{1-\left(\widehat{n}_{2_{x}}\cdot\widehat{n}_{3_{x}}\tan\widetilde{\theta}_{3}+\widehat{n}_{2_{y}}\cdot\widehat{n}_{3_{y}}\tan\widetilde{\theta}_{3}+\widehat{n}_{2_{z}}\cdot\widehat{n}_{3_{z}}\tan\widetilde{\theta}_{3}\right)\tan\widetilde{\theta}_{2}}
\end{split}
\label{eq:ERG-q-1}
\end{equation}
where $\widehat{n}_{3_{x,y,z}}\tan\widetilde{\theta}_{3}$ axial components are replaced by:
\begin{equation}
\begin{split}
\widehat{n}_{3_{x}}\tan\widetilde{\theta}_{3}&\mapsto\frac{\widehat{n}_{3_{x}}\tan\widetilde{\theta}_{3}
+\widehat{n}_{4_{x}}\tan\widetilde{\theta}_{4}-\left(\widehat{n}_{3_{z}}\cdot\widehat{n}_{4_{y}}\tan\widetilde{\theta}_{4}-\widehat{n}_{3_{y}}\cdot\widehat{n}_{4_{z}}\tan\widetilde{\theta}_{4}\right)\tan\widetilde{\theta}_{3}}
{1-\left(\widehat{n}_{3_{x}}\cdot\widehat{n}_{4_{x}}\tan\widetilde{\theta}_{4}+\widehat{n}_{3_{y}}\cdot\widehat{n}_{4_{y}}\tan\widetilde{\theta}_{4}+\widehat{n}_{3_{z}}\cdot\widehat{n}_{4_{z}}\tan\widetilde{\theta}_{4}\right)\tan\widetilde{\theta}_{3}}\\
\widehat{n}_{3_{y}}\tan\widetilde{\theta}_{3}&\mapsto\frac{\widehat{n}_{3_{y}}\tan\widetilde{\theta}_{3}
+\widehat{n}_{4_{y}}\tan\widetilde{\theta}_{4}-\left(\widehat{n}_{3_{x}}\cdot\widehat{n}_{4_{z}}\tan\widetilde{\theta}_{4}-\widehat{n}_{3_{z}}\cdot\widehat{n}_{4_{x}}\tan\widetilde{\theta}_{4}\right)\tan\widetilde{\theta}_{3}}
{1-\left(\widehat{n}_{3_{x}}\cdot\widehat{n}_{4_{x}}\tan\widetilde{\theta}_{4}+\widehat{n}_{3_{y}}\cdot\widehat{n}_{4_{y}}\tan\widetilde{\theta}_{4}+\widehat{n}_{3_{z}}\cdot\widehat{n}_{4_{z}}\tan\widetilde{\theta}_{4}\right)\tan\widetilde{\theta}_{3}}\\
\widehat{n}_{3_{z}}\tan\widetilde{\theta}_{3}&\mapsto\frac{\widehat{n}_{3_{z}}\tan\widetilde{\theta}_{3}
+\widehat{n}_{4_{z}}\tan\widetilde{\theta}_{4}-\left(\widehat{n}_{3_{y}}\cdot\widehat{n}_{4_{x}}\tan\widetilde{\theta}_{4}-\widehat{n}_{3_{x}}\cdot\widehat{n}_{4_{y}}\tan\widetilde{\theta}_{4}\right)\tan\widetilde{\theta}_{3}}
{1-\left(\widehat{n}_{3_{x}}\cdot\widehat{n}_{4_{x}}\tan\widetilde{\theta}_{4}+\widehat{n}_{3_{y}}\cdot\widehat{n}_{4_{y}}\tan\widetilde{\theta}_{4}+\widehat{n}_{3_{z}}\cdot\widehat{n}_{4_{z}}\tan\widetilde{\theta}_{4}\right)\tan\widetilde{\theta}_{3}}
\end{split}
\label{eq:ERG-q-2}
\end{equation}
\end{widetext}
Phase-shifts expressions $_{1'}^{4}\widetilde{\Phi}_{gg}$ and $_{4'}^{1}\widetilde{\Phi}_{gg}$ can thus be analytically obtained encapsulating Eq.~\ref{eq:ERG-p-1} with Eq.~\ref{eq:ERG-p-2} and Eq.~\ref{eq:ERG-q-1} with Eq.~\ref{eq:ERG-q-2} into Eq.~\ref{eq:GHR-2pulses-1} and Eq.~\ref{eq:GHR-2pulses-2}.
\begin{figure}[t!!]
\center
\resizebox{8.5cm}{!}{\includegraphics[angle=0]{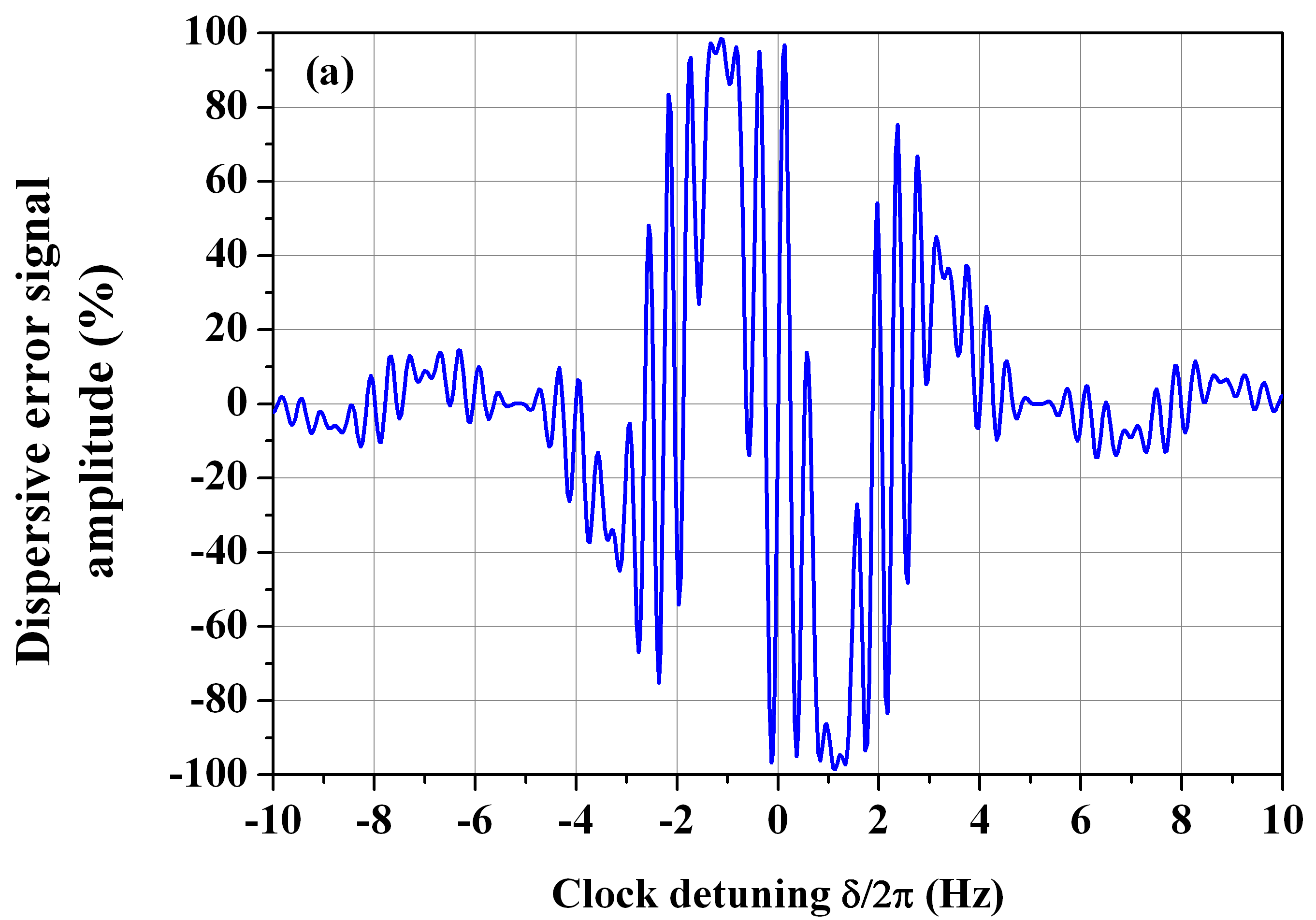}}
\resizebox{8.5cm}{!}{\includegraphics[angle=0]{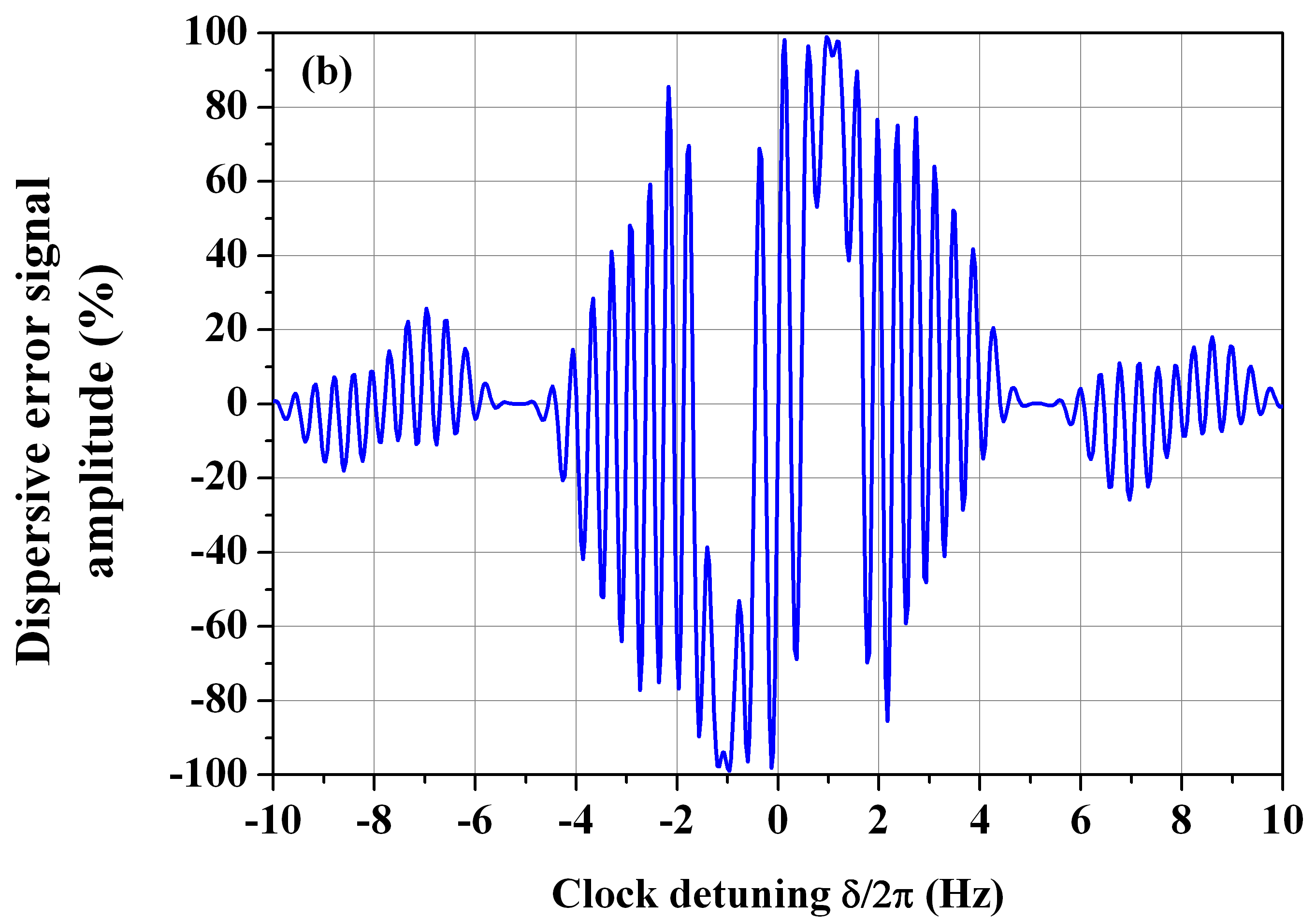}}
\caption{(color online). Examples of dispersive error signals based on $_{2'}^{2}C_{gg}$, calculated with Eq.~\ref{eq:Ramsey-transition} and applying the ERG transformation through Eq.~\ref{eq:new-coefficients}. (a) $90'^{\circ}_{\pm\pi/2}90'^{\circ}\dashv\delta\textup{T}\vdash270^{\circ}$, (b) $90'^{\circ}_{\pm\pi/2}90'^{\circ}\dashv\delta\textup{T}\vdash180_{\pi}^{\circ}90^{\circ}$. The Rabi frequency for all pulses is $\Omega=\pi/2\tau$ where the pulse duration reference is $\tau=3/16$~s, the free evolution time is T$=2$~s.}
\label{fig:spinor-interferences}
\end{figure}

\subsection*{C: spinor interferences with composite pulses}

\indent We apply our ERG algorithm following Eq.~\ref{eq:transformation-pq-pulses} to derive transition probabilities of composite spinor interferences based on Eq.~\ref{eq:transition-probability-Pauli}. We use a quantization axis projection with $\widehat{m}=(0,0,1)$ leading to $\widetilde{\theta}_{m_{z}}=\delta\textup{T}/2$.
We first derive the Ramsey formula with two pulses based on the coefficient $_{1'}^{1}C_{gg}$ as following:
\begin{equation}
\begin{split}
_{1'}^{1}C_{gg}&=_{1'}^{1}\widetilde{C}^{+}_{gg}e^{i_{1'}^{1}\widetilde{\Phi}_{gg}^{+}}e^{i\widetilde{\theta}_{m_{z}}}+_{1'}^{1}\widetilde{C}^{-}_{gg}e^{i_{1'}^{1}\widetilde{\Phi}_{gg}^{-}}e^{-i\widetilde{\theta}_{m_{z}}}\\
_{1'}^{1}\widetilde{C}_{gg}^{\pm}&=\frac{1}{2}\left(\cos\widetilde{\theta}_{1'}\cos\widetilde{\theta}_{1}\right)~_{1'}^{1}C^{\pm}\sqrt{1+\tan^{2}\left(_{1'}^{1}\widetilde{\Phi}_{gg}^{\pm}\right)}
\end{split}
\label{eq:Ramsey-transition}
\end{equation}
where phase-shifts are given by:
\begin{equation}
\begin{split}
_{1'}^{1}\widetilde{\Phi}_{gg}^{+}=&\arctan\left[\frac{\widehat{n}_{1'_{z}}\tan\widetilde{\theta}_{1'}+\widehat{n}_{1_{z}}\tan\widetilde{\theta}_{1}}{1-\widehat{n}_{1'_{z}}\widehat{n}_{1_{z}}\tan\widetilde{\theta}_{1'}\tan\widetilde{\theta}_{1}}\right]\\
_{1'}^{1}\widetilde{\Phi}_{gg}^{-}=&\arctan\left[\frac{\left(\widehat{n}_{1'_{y}}\widehat{n}_{1_{x}}-\widehat{n}_{1'_{x}}\widehat{n}_{1_{y}}\right)\tan\widetilde{\theta}_{1'}\tan\widetilde{\theta}_{1}}{\left(\widehat{n}_{1'_{x}}\widehat{n}_{1_{x}}+\widehat{n}_{1'_{y}}\widehat{n}_{1_{y}}\right)\tan\widetilde{\theta}_{1'}\tan\widetilde{\theta}_{1}}\right]
\end{split}
\label{eq:Ramsey-phase-shift}
\end{equation}
and:
\begin{equation}
\begin{split}
_{1'}^{1}C^{+}&=2-2\widehat{n}_{1'_{z}}\tan\widetilde{\theta}_{1'}\cdot\widehat{n}_{1_{z}}\tan\widetilde{\theta}_{1}\\
_{1'}^{1}C^{-}&=-2\widehat{n}_{1'_{x}}\tan\widetilde{\theta}_{1'}\cdot\widehat{n}_{1_{x}}\tan\widetilde{\theta}_{1}-2\widehat{n}_{1'_{y}}\tan\widetilde{\theta}_{1'}\cdot\widehat{n}_{1_{y}}\tan\widetilde{\theta}_{1}
\end{split}
\label{eq:Ramsey-C-coefficients}
\end{equation}
We now proceed with our ERG algorithm to obtain coefficients of the $_{2'}^{2}C_{gg}$ amplitude of transition using $p=2'$ pulses on the left arm and $q=2$ pulses on the right arm of the spectroscopic pulse scheme. We get:
\begin{equation}
\begin{split}
\cos\widetilde{\theta}_{1'}&\mapsto\cos\widetilde{\theta}_{1'}\cos\widetilde{\theta}_{2'}\left(1-\widehat{n}_{1'}\tan\widetilde{\theta}_{1'}\cdot\widehat{n}_{2'}\tan\widetilde{\theta}_{2'}\right)\\
\cos\widetilde{\theta}_{1}&\mapsto\cos\widetilde{\theta}_{1}\cos\widetilde{\theta}_{2}\left(1-\widehat{n}_{1}\tan\widetilde{\theta}_{1}\cdot\widehat{n}_{2}\tan\widetilde{\theta}_{2}\right)\\
\widehat{n}_{1'_{x,y,z}}\tan\widetilde{\theta}_{1'}&\mapsto\left(\right)^{2'}_{x,y,z}\\
\widehat{n}_{1_{x,y,z}}\tan\widetilde{\theta}_{1}&\mapsto\left(\right)^{2}_{x,y,z}\\
_{1'}^{1}\widetilde{\Phi}_{gg}^{\pm}&\mapsto_{2'}^{2}\widetilde{\Phi}_{gg}^{\pm}
\end{split}
\label{eq:new-coefficients}
\end{equation}
where modified elements $\left(\right)^{2'}_{x,y,z}$ and $\left(\right)^{2}_{x,y,z}$ are given by Eq.~\ref{eq:GHR-2pulses-1} and Eq.~\ref{eq:GHR-2pulses-2}. We have plotted two examples of arbitrary composite spinor interferences based on four pulses versus the clock detuning in Fig.~\ref{fig:spinor-interferences}.



\begin{thebibliography}{0}
\bibitem{Note} A second recursive algorithm related to a M\"{o}bius transformation in conformal mapping is based on the following modification with $l\in\{p,q\}$ $(p\rightarrow+,q\rightarrow-)$ as:
\begin{equation*}
\begin{split}
\tan\widetilde{\theta}_{l}\mapsto\frac{\tan\widetilde{\theta}_{l}+e^{-i\Xi_{l,l+1}}\tan\widetilde{\theta}_{l+1}}{1-e^{-i\Xi_{l,l+1}}\tan\widetilde{\theta}_{l}\tan\widetilde{\theta}_{l+1}}
\end{split}
\label{}
\end{equation*}
where $\Xi_{\textup{l,l+1}}=\phi_{l}+\phi_{l+1}\pm(\varphi_{l}-\varphi_{l+1})$.
\end{thebibliography}
\end{document}